\documentclass[pre,aps,superscriptaddress,onecolumn]{revtex4-1}
\usepackage{graphicx} 
\graphicspath{{graph/}}
\usepackage[usenames]{color}
\usepackage{amsmath,amssymb}
\usepackage{gensymb}
\usepackage{natbib}
\usepackage{xcolor}
\def\strutdepth{\dp\strutbox}
\def\nw#1{\strut\vadjust{\kern-\strutdepth\vtop to0pt{\vss\hbox to\hsize
{\hskip\hsize\hskip5pt$\leftarrow$\hss\strut}}}{\em #1}}

\usepackage{rotating}
\usepackage{multirow}

\begin{document}


\title{Spreading on viscoelastic solids: Are contact angles selected by Neumann's law?}

\author{M. van Gorcum}
\affiliation{Physics of Fluids Group, Faculty of Science and Technology, Mesa+ Institute, University of Twente, 7500 AE Enschede, The Netherlands.}
\author{S. Karpitschka}
\affiliation{Max Planck Institute for Dynamics and Self-Organization (MPIDS), 37077 Göttingen, Germany} 
\author{B. Andreotti}
\affiliation{Laboratoire de Physique Statistique, UMR 8550 ENS-CNRS, Univ. Paris-Diderot, 24 rue Lhomond, 75005, Paris.}
\author{J. H. Snoeijer}
\affiliation{Physics of Fluids Group, Faculty of Science and Technology, Mesa+ Institute, University of Twente, 7500 AE Enschede, The Netherlands.}

\date{\today}

\begin{abstract}
The spreading of liquid drops on soft substrates is extremely slow, owing to strong viscoelastic dissipation inside the solid. A detailed understanding of the spreading dynamics has remained elusive, partly owing to the difficulty in quantifying the strong viscoelastic deformations below the contact line that determine the shape of moving wetting ridges. Here we present direct experimental visualisations of the dynamic wetting ridge, complemented with measurements of the liquid contact angle. It is observed that the wetting ridge exhibits a rotation that follows exactly the dynamic liquid contact angle -- as was previously hypothesized [Karpitschka \emph{et al.} Nature Communications \textbf{6}, 7891 (2015)]. This experimentally proves that, despite the contact line motion, the wetting ridge is still governed by Neumann's law. Furthermore, our experiments suggest that moving contact lines lead to a variable surface tension of the substrate. We therefore set up a new theory that incorporates the influence of surface strain, for the first time including the so-called Shuttleworth effect into the dynamical theory for soft wetting. It includes a detailed analysis of the boundary conditions at the contact line, complemented by a dissipation analysis, which shows, again, the validity of Neumann's balance.
\end{abstract}


\maketitle


\section{Introduction}
The interfacial properties of liquids and polymeric solids are fundamental to nanometer scale devices, with applications in tribology and lubrication, transport across membranes, nanofluidic devices, and biological systems. However, it has remained a challenge to characterise the interfacial mechanics of soft solids~\cite{style2017elastocapillarity,NadermannPNAS2013,MondalPNAS2015,Andreotti2016a,AndreottiARFM2020}. It has been proposed recently that liquid drops can serve as an effective tool to quantify the interfacial mechanics~\cite{xu2017direct,schulman2018surface,snoeijer2018paradox,xu2018}. Namely, droplets act on the solid with an extremely localised traction, probing the nanoscale, since molecular interactions are localised over the thickness of the interface. The droplets deform the soft solid into a ``wetting ridge" that moves along with the contact line. Besides capillary forces, contact line motion therefore also probes the viscoelastic response of the polymer \cite{shanahan1994anomalous,Carre1996a,LAL96,LALLang96,KarpNcom15,zhao2018geometrical}. For example, it has been shown that the response of nanometer scale polymers grafted or adsorbed at a surface presents a scaling law consistent with the picture emerging from statistical physics~\cite{Lhermerout2016aa}. However, in order for droplets to be fully useful as a quantitative rheological tool, one must have a perfect theoretical understanding of the processes at work. 

Spreading drops and contact line motion have been extensively studied on rigid surfaces~ \cite{deGe02,RevBonn09,RevSnoB13}. Detailed hydrodynamic analysis has demonstrated how the liquid interface is affected by contact line motion~\cite{V76,Cox86,tanner1979spreading}. This leads to dynamic (macroscopic) contact angles, which differ from the equilibrium angles and which depend on the contact line velocity. The intricate mechanics have been compactly summarized via a dissipation analysis~\cite{dG85,deGe02}, balancing the power injected by capillary forces with the dissipation in the vicinity of the contact line. Mechanical and dissipation approaches were shown to be strictly equivalent~\cite{RevBonn09}, though the expressions for the dynamic contact angle found in the literature can appear slightly different due to different levels of mathematical approximations.

The spreading of drops over soft surfaces was first addressed in a series of papers by Carr\'e \& Shanahan~\cite{Shanahan1995aa,shanahan1994anomalous,Carre1996a,Shanahan2002a}, and by Long, Ajdari \& Leibler~\cite{LAL96,LALLang96}. The main observation is that contact line motion is slowed down dramatically as compared to spreading over rigid surfaces. This slowing down can be attributed to the strong dissipation in the polymer layer, and was termed ``viscoelastic braking". From a modelling perspective, the dynamic contact angles were estimated using a dissipation approach~\cite{Carre1996a,LALLang96}. 
\begin{figure}[tb!]
\centering
\includegraphics[width=86mm]{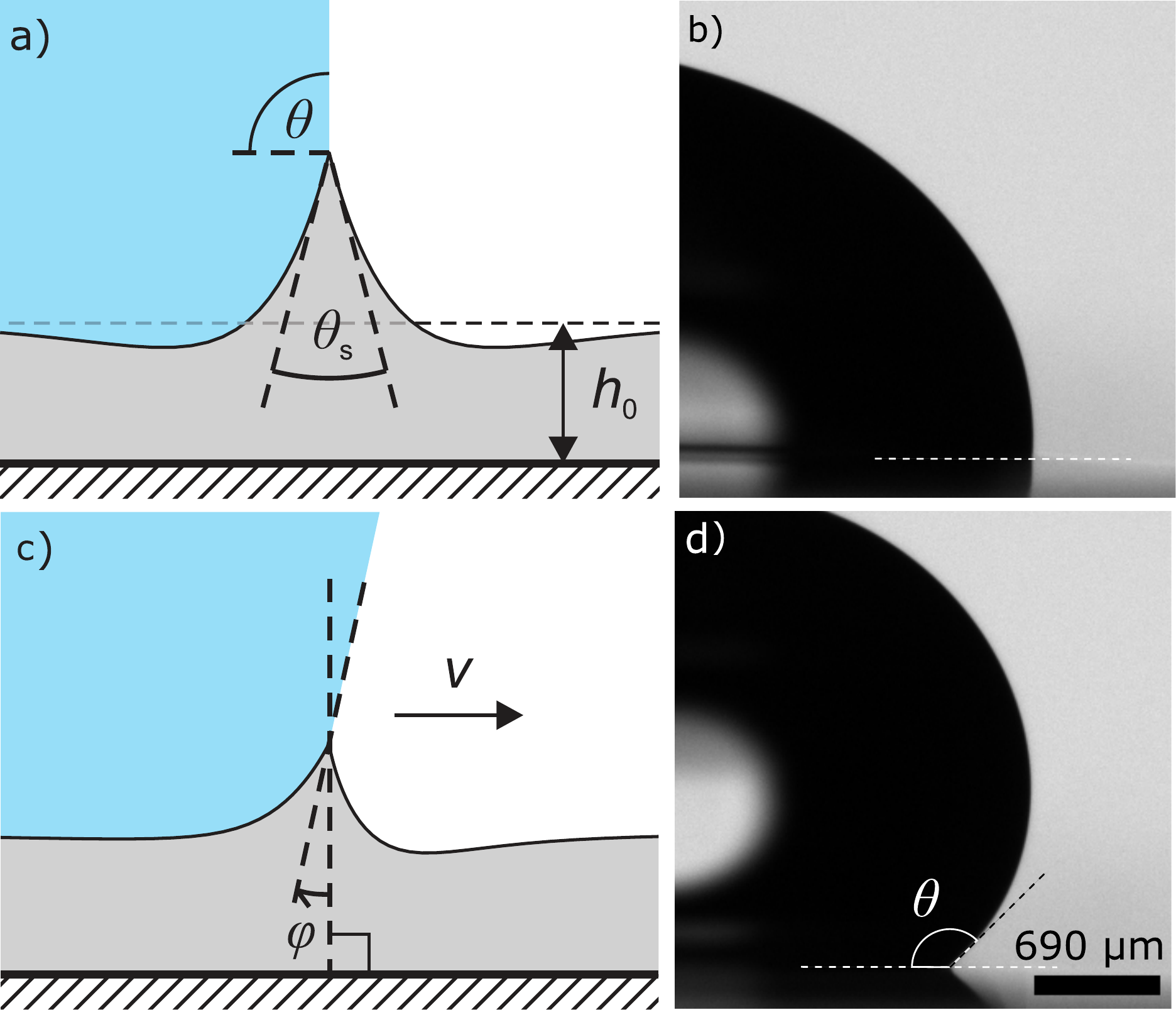}
  \caption{(ab) Soft wetting at equilibrium. (a)~Zoom of the wetting ridge near the contact line on the scale of the elastocapillary length $\gamma/G$, and the definition of the liquid angle $\theta$ and the solid angle $\theta_S$. The profile is computed from linear theory. (b)~Experimental side view image of a static drop. (cd) Soft wetting dynamics. (c)~When the contact line is moving, the viscoelasticity of the substrate leads to a rotation of the wetting ridge by an angle $\Delta \theta$ (again computed from linear theory). (d) Experimental side view image of a spreading drop, which exhibits a dynamic contact angle $\theta$.}
  \label{fgr:figure1}
\end{figure}

In recent years there have been major advances on the wetting of soft substrates~\cite{style2017elastocapillarity,Andreotti2016a,Bico2018ARFM,AndreottiARFM2020}. A variety of experimental methods have provided detailed information on the wetting ridge below the contact line~\cite{PC2008aa,PARKNATURE,Jerison2011a,Style13,PRLstick-slip}, complemented by theoretical developments~\cite{Limat2012a,Style2012a,MDSA12b,Bostwick:2014aa,Lub14,snoeijer2018paradox,Masurel2019}. The typical size of the wetting ridge is given by the ratio of surface tension of the drop $\gamma$ and the substrate's shear modulus $G$, which defines the elastocapillary length $\gamma/G$. A striking feature is that, at equilibrium, the ridge satisfies the Neumann law: similar to liquid interfaces, the surface tension of the solid balances the traction imposed by drop. This gives rise to a well-defined solid angle, defined as $\theta_S$ in figure~\ref{fgr:figure1}a. 

While at equilibrium Neumann's law can be derived from energy minimisation~\cite{Lub14,snoeijer2018paradox}, there is no consensus as to whether it is valid for wetting dynamics~\cite{Karpitschka2018,zhao2018geometrical}. In a previous study~\cite{KarpNcom15} we hypothesised that the dynamic liquid angle is selected by a rotation of the wetting ridge, while maintaining the Neumann angles. This mechanism is sketched in figure~\ref{fgr:figure1}cd: the motion of the contact line induces a rotation $\varphi$ which is followed by a change in the liquid angle $\Delta \theta =\theta-\theta_{eq}$. In case the Neumann law applies, one thus finds $\Delta \theta = \varphi$. However, this point of view was challenged, claiming that the dissipation-based theory implies that Neumann's law is not valid in dynamical situations~\cite{zhao2018geometrical} due to the appearance of a perfectly localised viscoelastic contribution that is able to compete with surface forces~\cite{Roche:arxiv19}.

An additional complexity to the problem, which arises even under static conditions, is that the surface tension of a solid interface cannot be assumed to remain constant. Owing to the so-called Shuttleworth effect~\cite{Shuttleworth1950a,Andreotti2016a,style2017elastocapillarity,AndreottiARFM2020}, the surface energy depends on the amount of surface strain. At equilibrium this strain dependence was recently confirmed \cite{xu2017direct,schulman2018surface,snoeijer2018paradox}, giving rise to variations of the solid angle $\theta_S$. Similar variations of the ridge geometry have been reported in dynamical experiments \cite{park2017self,PRLstick-slip}, though a systematic experimental observation of all the contact angles is still lacking. 

The aim of this paper is to address a series of unresolved issues, which naturally emerge from these recent experimental and theoretical developments. These are centred around question on how the contact angles are selected during the spreading of drops over viscoelastic substrates: 

\begin{itemize}
\item Is Neumann's law still applicable for moving contact lines? 
\item Is the change of the liquid angle directly associated with a rotation of the wetting ridge? 
\item To what extent can these relations be derived from a power balance or from a stress balance, which obviously must lead to the same answer?
\item Finally, how is the dynamics affected by the Shuttleworth effect? 
\end{itemize}

The paper starts by an experimental quantification of the dynamic contact angles from direct visualisation of moving wetting ridges. The experimental method is described in Sec.~\ref{sec:exp}, while the results are presented in Sec.~\ref{sec:results}. At low velocity, we find a perfect agreement between the independently measured solid rotation $\varphi$ and the change of the liquid angle $\Delta \theta$: the equality $\Delta \theta= \varphi$ provides direct experimental evidence that Neumann's law can be applied for moving wetting ridges. However, at larger speeds, we also observe a change in $\theta_S$, which in a visco-elasto-capillary continuum formulation can only arise through a variable surface tension. In Sec.~\ref{sec:theory} we therefore setup a systematic route to solving the fully nonlinear problem including the Shuttleworth effect, which we work out to lowest order in Sec.~\ref{sec:green} using a Green's function formalism. The paper closes with a critical discussion in Sec.~\ref{sec:discussion}, summarising the main open issues.

\section{Experimental set-up}\label{sec:exp}

We measure the dynamical shape of wetting ridges formed by moving contact lines. Our specific aim here  is to determine the angles that describe the local geometry of the three-phase region: the liquid angle $\theta$, the solid angle $\theta_S$, and rotation of the ridge $\Delta \theta$ (see Fig.~\ref{fgr:figure1}). Due to the topography of the wetting ridge it is challenging to resolve the solid angles with sufficient accuracy. Here we design an experimental setup that allows for a direct visualisation of the wetting ridge with unprecedented spatio-temporal resolution (see Sec.~\ref{subsect:cavity}). The liquid angle is measured in a separate experiment using a classical drop-on-planar-substrate geometry (Sec.~\ref{subsect:liquidangle}).

\begin{figure*}[tb]
\centering
	\includegraphics{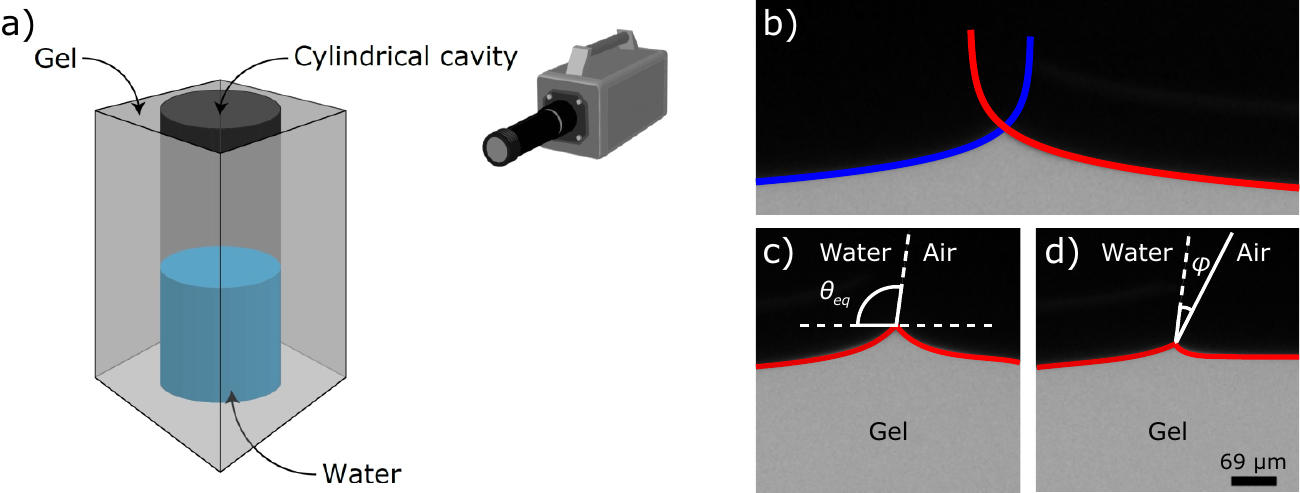}
  \caption{ (a) A rectangular cuvette filled with transparent gel with a cylindrical cavity is observed perpendicular to the sidewall as the cavity is filled with water from the bottom. A backlight illuminates the cuvette through a diffuser plate. (b) Logarithmic fits of the gel shape on either sides of the contact line are used to find the ridge tip and the contact angles. (c) Example of a static wetting ridge, and (d) of a dynamic ridge from which we determine the rotation angle $\Delta \theta$ (right). In the latter case, the contact line moves to the right.}
  \label{fgr:schematic_setup}
\end{figure*}

\subsection{Visualizing the wetting ridge}\label{subsect:cavity}

The visualization of the wetting ridge is performed with the setup sketched in figure \ref{fgr:schematic_setup}a. The idea is to create a nearly cylindrical cavity inside a square block of a soft viscoelastic substrate material. The cavity can then be filled partially with water, creating a single moving contact line that exerts a capillary traction onto the substrate. This traction points toward the center of the cavity and deforms its surface into an axisymmetric wetting ridge with a cross-section that is virtually identical to a wetting ridge on a planar surface (as long as the radius of the cavity is much larger than the ridge). In this configuration the wetting ridge can be imaged shadowgraphically through the planar faces of the square block, thus minimizing any optical distortions. 

The liquid used in the experiments is deionized water. For the polymer gel we have chosen two different reticulated polymer networks: a polydimethylsiloxane (PDMS) gel (Dow Corning CY52-276 mixed at a 1.3:1 (A:B) ratio), and a polyvinyl siloxane (PVS) gel (Esprit Composite RTV EC00 mixed at 1:2.5 (base:catalyst) ratio). Both are referred to as gels,  in the sense that they cross a gelation transition during curing, at which the system presents a vanishing shear modulus and a diverging viscosity at low frequencies. Both gels are prepared such that their static shear modulus after curing is around $G=400$~Pa. The viscoelastic rheology is accurately fitted by a simple power-law form

\begin{eqnarray}\label{eq:gel1}
\mu(\omega) = G'(\omega) + iG''(\omega) = G\left[1+(i\omega \tau)^n\right],
\end{eqnarray}
where for details we referent to we refer to Appendix~\ref{app:rheology}. The (static) elastocapillary length is then around $\gamma/G=180 \, \mu$m, and leads to relatively large wetting ridges that are comfortably measurable. 

The cylindrical cavity is created by the following procedure. We first fill a standard spectroscopy cuvette (with inner dimensions of 1 x 1 x 4.5 cm) with uncured but mixed and degassed liquid components of the gel, leaving a small air volume at the top and seal the open end of the cuvette. The cuvette is then spun at $\approx100$ RPS about its long axis, such that the centrifugal forces turn the air volume into nearly cylindrical cavity extending to the bottom of the cuvette. We verify that the radius of the cavity (typically 4 mm) is constant within the measurement section ($\sim 2$ mm). The diameter of the cavity and the gel thickness are much larger than the elastocapillary length ($\gamma/G$), while the Bond number ($\Delta\rho G L^2/\gamma$) remains low to keep the effect of gravity negligible. The gel is cured while spinning at room temperature for $\sim 14$ hours. For the PDMS gel, the cuvette is additionally heat cured afterwards in an oven at 80 degrees for two hours.

The gel surface is observed using a long distance video microscope perpendicular to the cuvette wall, focused on the diametral plane of the cavity. The cuvette is illuminated with diffuse light from the back (figure~\ref{fgr:schematic_setup}). The cavity reflects and refracts light while the bulk gel is transparent. Thus the gel appears bright, and the cavity dark. Then the cavity is partly filled with MilliQ water. The capillary deformation caused by the water meniscus can be directly observed, with a spatiotemporal resolution limited only by the optical properties of the shadowgraphy setup. In these experiments a 2-4x lens is used, leading to a pixel scale of $\approx 2\mu$m per pixel and a field of view of about $2.2 \times 1.6$~mm$^2$. The images were recorded with a CMOS camera at rates of 2 to 52 frames per second, and with a high-speed camera (Photron FASTCAM Mini UX100) at frame rates between $50$ and $3200$~fps.

The gel interface profile is detected with sub-pixel accuracy by fitting the greyscale profile in the vertical direction by an error function, locating the interface at its inflection point. The tip of the wetting ridge is rounded in our measurement due to the diffraction limit of the shadowgrapy setup. The typical radus of curvature detected for the blurred image of the ridge tip is about $2-3$ $\mu$m, independent of the imaging scale. The ridge is found to be sharp down to the optical resolution, with a well defined opening angle $\theta_S$, as observed previously by X-ray microscopy \cite{park2017self, PARKNATURE}. While the PDMS gel is optically clear, the PVS gel is slightly opaque, resulting in a reduced contrast of the wetting ridge. Within our subpixel resolution scheme, this can be partially corrected for, yet the measurements on the PVS gel have a slightly lower precision.

To extract the relevant angles we extrapolate the surface profiles from both sides into the diffraction limited region at the ridge tip. Because the elastic response to a point force is known to be logarithmic, we use a least squares fit of a generic logarithm function $f=a+b \log(\pm (x-c_{\pm}))$ to the left (-) and right (+) of the contact line, as shown in figure~\ref{fgr:schematic_setup}b.
$c_{\pm}$ does not coincide with the contact line location, which accounts for the elastocapillary effects near the contact line. For $x\gg c$, the function converges to the expected logarithmic shape. Down to the resolution limit, there was no systematic deviation from the experimental data detectable. The intersection of the extrapolated fits is used to measure the solid opening angle ($\theta_S$) and the relative rotation angle ($\varphi$). The latter is defined as the angle between the horizontal and the bisector of the two profile fits (figure~\ref{fgr:schematic_setup}c). As long as the change in solid opening angle is small, or the change in $\theta_S$ is symmetric for the liquid and vapor-sides of the gel, this gives an accurate measurement of the ridge rotation $\varphi$ (we anticipate that this is no longer the case at large velocity). A linear regression on the horizontal position of the intersection is used to measure the contact line speed.

The experiments using 2 fps allow us to measure very slow dynamics, where the experiments were run for approximately 10 minutes. The gradual deceleration of the contact line allowed us to resolve velocities down to $\sim 1$ nm/s. In the high-speed measurements, we could resolve the fast dynamics at the depinning transition of the contact line down to sub-millisecond temporal resolution.

\subsection{Liquid contact angle}\label{subsect:liquidangle}

The liquid contact angle measurements are performed with a DataPhysics OCA 15 apparatus. A $\sim 10\; \mu$L drop is deposited on a $12$~mm thick gel layer and the droplet volume is quickly increased to $\sim 20\;\mu$L. After this increase, the droplet relaxes toward its equilibrium wetting configuration. All measurements presented here are performed during this overdamped relaxation phase, governed by the dissipation in the viscoelastic substrate. The liquid contact angle $\theta$ and contact line speed~$v$ are determined from the recorded video by using a sub-pixel resolution edge detection in MATLAB.
The edge detection scheme uses a threshold value for the pixel scale edge detection and a linear interpolation around the edge to find the sub-pixel position. The speed is deduced from the variation of the contact line position, with a resolution down to $\sim 1$ nm/s. To obtain $\Delta \theta = \theta - \theta_{eq}$, the contact angle $\theta_{eq}$ was determined at vanishing speed.

\section{Experimental results}\label{sec:results}

\subsection{Phenomenology}

Before coming to detailed measurements of the contact angles, let us first describe the types of wetting dynamics that are observed in our system, and relate this to previous observations in the literature.

\begin{figure}[tb]
\centering
  \includegraphics[width=86mm]{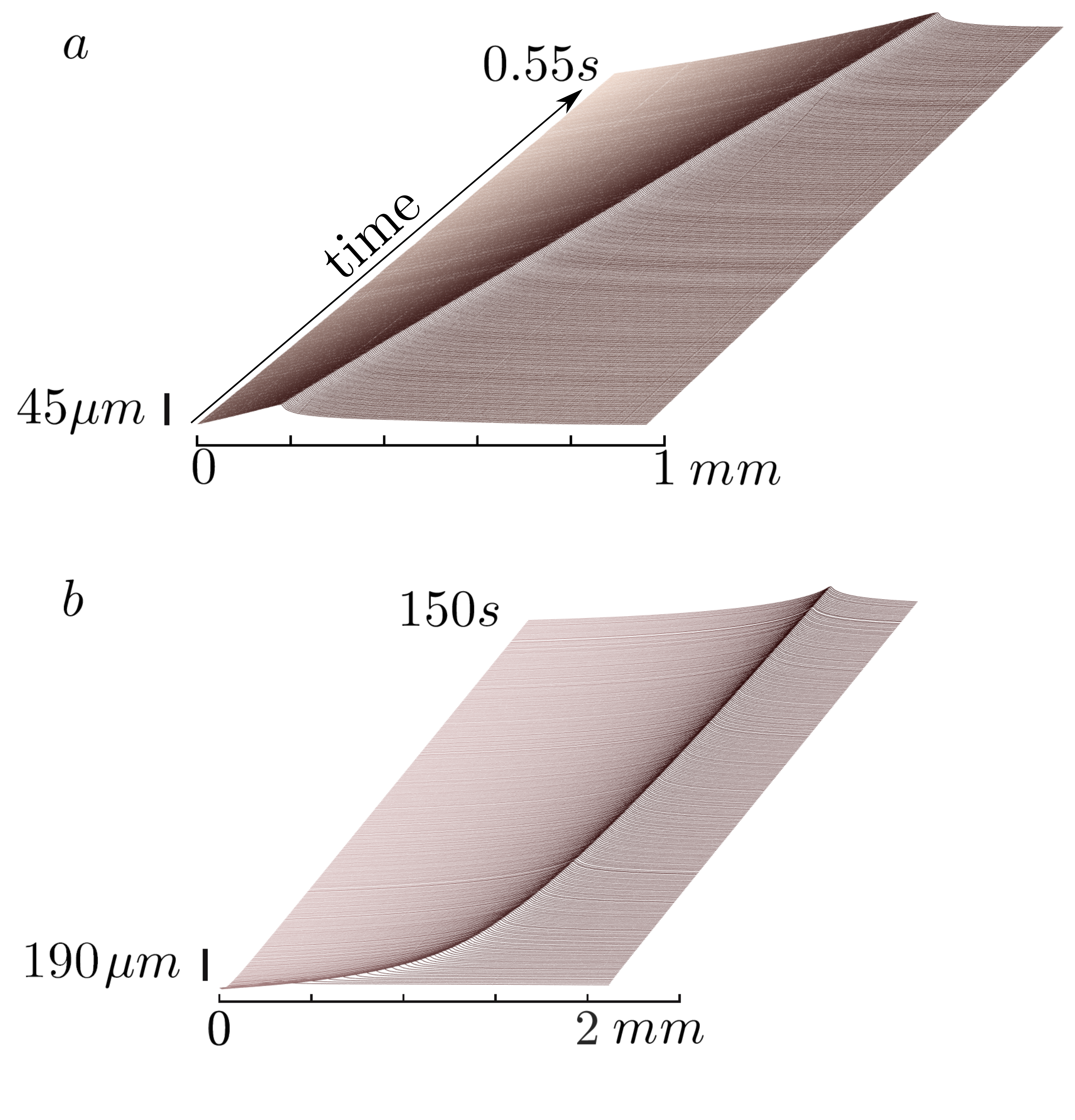}
  \caption{Space-time diagrams for slow dynamics. (a) Wetting ridge moving at a constant velocity, for $v<v_c$. (b) Relaxation of the wetting ridge towards a static equilibrium, after stopping the injection of water into the cavity. The size of the wetting ridge is indicated by the scale bars. The static ridge is significantly larger than the dynamic ridge.}  
  \label{fgr:spacetimes_steady}
\end{figure}

In figure~\ref{fgr:spacetimes_steady}a we show a typical experimental result in the ``slow" regime, where water is injected into the cavity at a constant, moderate flow rate. The image is obtained by stacking the gel profiles at different times, so that one can track the temporal evolution of the wetting ridge. In the case of figure~\ref{fgr:spacetimes_steady}a, the contact line moves at a constant velocity $v$ such that the ridge tip location imprints a straight line to the space-time plot. For this steady motion, the liquid contact angle and the shape of the moving liquid meniscus are stationary in the comoving frame. Thus mass conservation dictates that the contact line velocity $v=Q/A$ where $Q$ is the imposed flow rate and $A$ is the cross-sectional area of the cavity. The contact angle $\theta$ takes on a value larger than $\theta_{eq}$ and depends on the (imposed) contact line velocity. 

\begin{figure}[tb]
\centering
  \includegraphics[width=86mm]{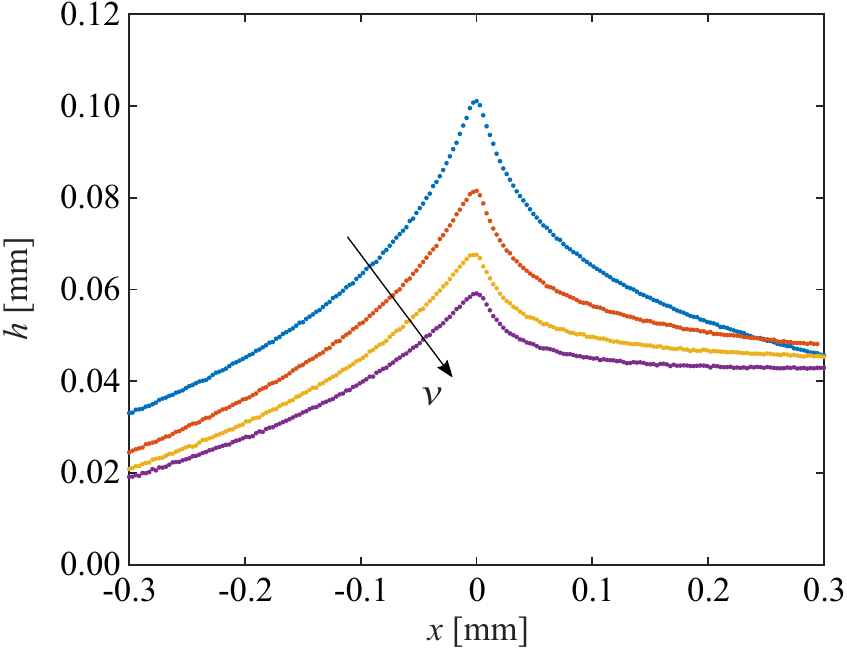}
  \caption{Experimental ridge shapes, for gradually increasing contact line speeds ($0$, $6.3\times10^{-3}$, $3.2\times10^{-2}$, $1.0\times10^{-1}$ mm/s). The contact line motion is towards the right.}
  \label{fgr:ridge_shapes}
\end{figure}

Then, at some time we suddenly stop the injection of fluid ($Q=0$) so that the contact line will eventually come to rest. This relaxation dynamics is shown in figure~\ref{fgr:spacetimes_steady}b. The wetting ridge does not stop instantaneously but relaxes over about 50-100 seconds since the contact angle $\theta > \theta_{eq}$. However, at the end of the sequence shown in figure~\ref{fgr:spacetimes_steady}b, the contact line velocity is still $\sim 10^{-3}$~mm/s, and it takes on the order of 5 minutes to reach an equilibrium position as far as is detectable within the measurement precision. This very slow relaxation is due to the strong viscoelastic dissipation associated to the contact line motion, an effect that was established already by Carr\'e \& Shanahan~\cite{Shanahan1995aa,shanahan1994anomalous,Carre1996a,Shanahan2002a} and Long, Ajdari \& Leibler~\cite{LAL96,LALLang96} who termed it ``viscoelastic braking". As can be seen in figure~\ref{fgr:spacetimes_steady}b, the wetting ridge also increases in amplitude during the relaxation. This is due to the fact that moving wetting ridges are smaller in amplitude than static ridges -- a fact that can be attributed to the frequency-dependence of the complex gel modulus $|\mu|$, which results in a ``dynamical elastocapillary length" that is decreasing with velocity.
Note the  scale bars indicated in both figure~\ref{fgr:spacetimes_steady}ab, showing that the wetting ridge is indeed relatively large in the static regime. More detailed zooms of dynamical profiles are given in figure~\ref{fgr:ridge_shapes}. Importantly, the dynamics in this relaxation regime is ``quasi-steady": we have verified that the contact angle $\theta$ depends on the instantaneous velocity $v$ in the same way as in the steady-state regime. Since dissipation governs the dynamic angle and is determined by both ridge size and velocity, this indicates that also the ridge size behaves quasi-steadily during deceleration and is an instantaneous function of $v$.

\begin{figure}[tb!]
\centering
  \includegraphics[width=86mm]{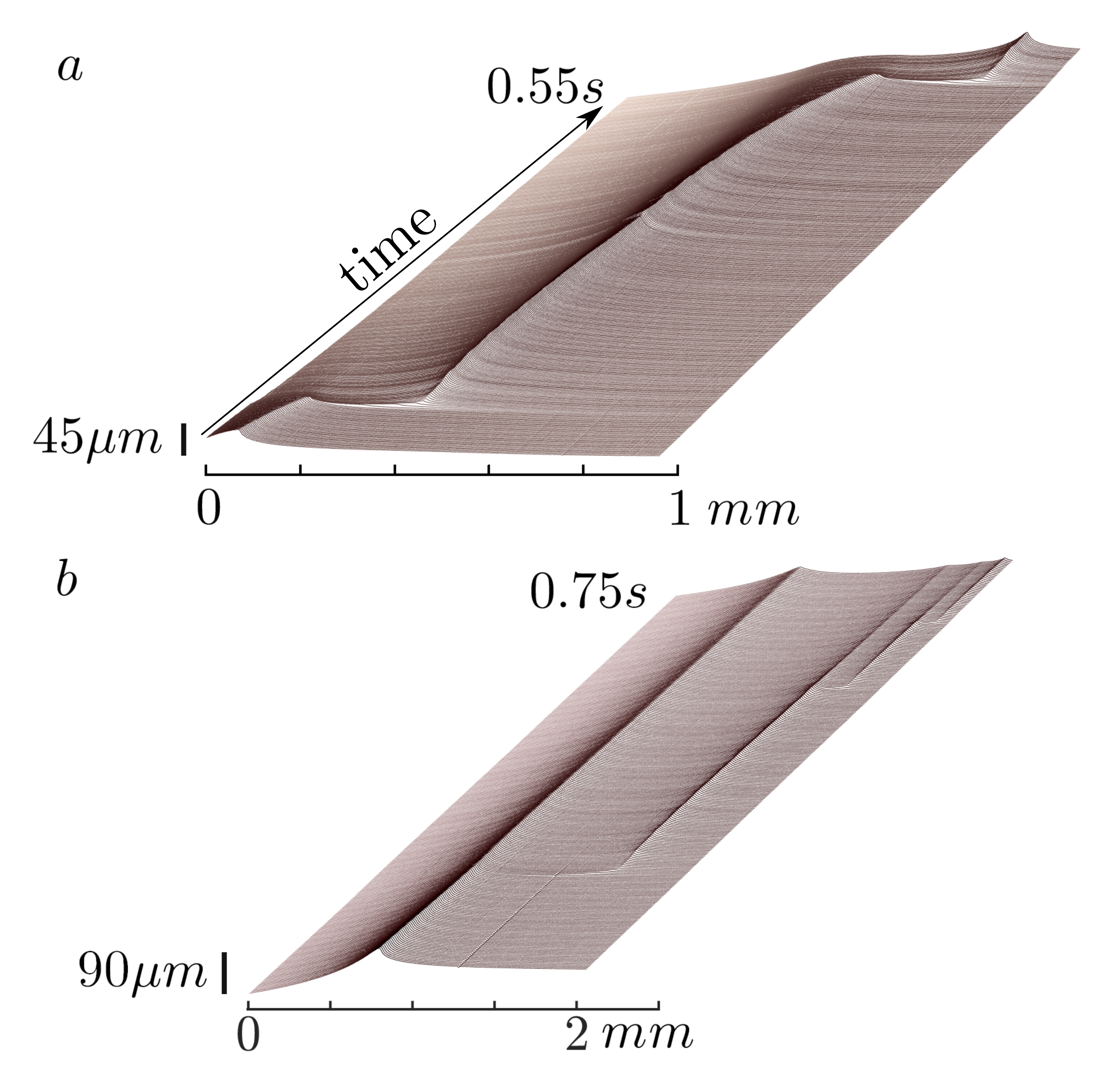}
  \caption{Space-time diagrams for rapid dynamics, leading to stick-slip motion. (a) Wetting ridge moving at high velocity, with $U>U_c$. Even though the imposed volume flux of water $Q$ into the cavity is constant, the contact line undergoes a stick-slip motion where it periodically ``depins" from its own wetting ridge. Here one cycle is shown. (b) Initial contact line motion, after starting the injection of liquid from an equilibrium state. The contact line first depins from a large static wetting ridge (which very slowly decays over time), leading to subsequent stick-slip motion. The size of the wetting ridge is indicated by the scale bars. The static ridge is significantly larger than the dynamic ridge.}
  \label{fgr:spacetimes_unsteady}
\end{figure}

The same setup can be also be used to quantify stick-slip motion of the contact line, which recently gained much attention \cite{PRLstick-slip, Kajiya2013a, Kajiya2014aa, park2017self}. Again, we first consider the situation where we impose a constant flow rate $Q$, but now at a sufficiently large value that $v=Q/A$ exceeds a critical velocity $v_c \sim 1$~mm/s for the PDMS gel. The system then undergoes a transition to stick-slip dynamics, a cycle of which is shown in figure~\ref{fgr:spacetimes_unsteady}a.

We recently characterised the stick-slip transition in detail~\cite{PRLstick-slip}, and identified the criterion for spontaneous depinning of the contact line from its own wetting ridge. It turned out that the depinning occurs due to a violation of the classical Gibbs inequality for sharp edges -- but in the case of soft, dynamic ridges, the angles of the ridge are intricate functions of the velocity. For details we refer to Ref.~\cite{PRLstick-slip}. After depinning, the rapid contact line motion slows down over the timescale of about $5$~ms, which is due to the growth of a new wetting ridge. After deceleration, the cycle is repeated. The typical velocities during the rapid slip are between $10^2$ and $10^3$~mm/s, which are comparable to contact line velocities on rigid surfaces~\cite{RevBonn09,RevSnoB13}. The stick-slip dynamics was also revealed by X-ray measurements \cite{park2017self}, who referred to this regime as ``stick-slipping by a medium sized ridge".

Finally we also visualize the onset of contact line motion, starting from an equilibrium wetting configuration and then suddenly impose a finite $Q$. The dynamics is shown in figure~\ref{fgr:spacetimes_unsteady}b. Since we start from rest, the initial wetting ridge is relatively large and the contact line does not immediately leave the wetting ridge. Instead, this initially causes a slow transient response of the ridge. However, as soon as the Gibb's inequality is violated, the contact line is able to depin from the initial ridge. This regime was referred to by \cite{park2017self} as ``stick-breaking by a fully grown ridge". After depinning, this fully developed static ridge is observed to decay very slowly, much slower than the dynamic ridges. This could point to a poroelastic contribution of uncrosslinked polymer chains, as was discussed also in \cite{zhao2018growth,berman2019singular}. After depinning, the rapid contact line motion undergoes a series of stick-slip cycles, similar to those reported in figure~\ref{fgr:spacetimes_unsteady}a.

\subsection{Contact angles}

We now turn to the main experimental results of this paper, where we quantify the contact angles of dynamical wetting ridges. These are extracted from the profiles at different velocities such as in figure~\ref{fgr:ridge_shapes}. Details of how the contact angles are determined are given in Sec.~\ref{sec:exp}. We restrict ourselves to the quasi-steady dynamics, for which the contact line does not depin from its own ridge. 

\subsubsection{Low velocity.~~~}
\begin{figure}[tb!]
\centering
  \includegraphics{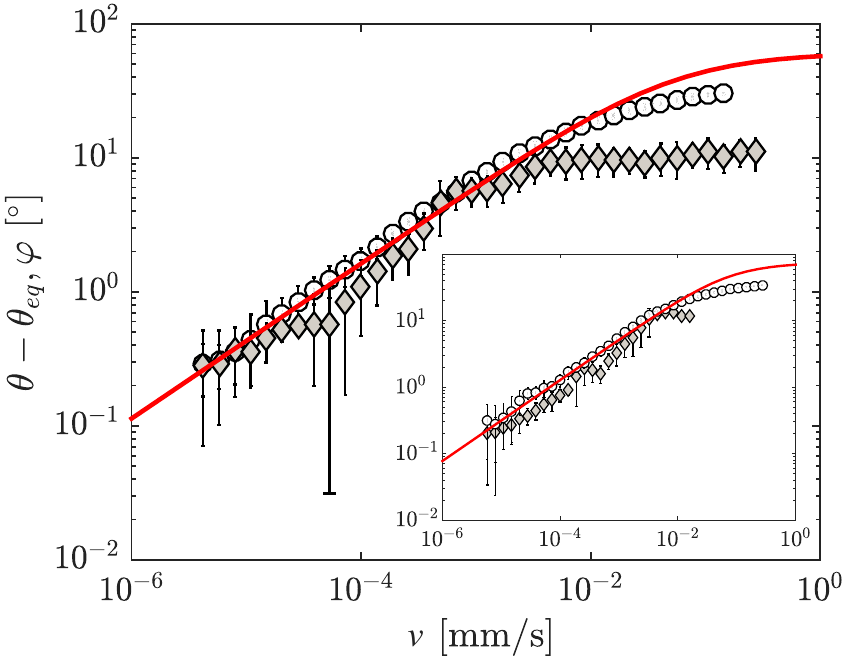} 
  \caption{Equality of the dynamic contact angle of the liquid and the ridge rotation angle. The main panel reports data on the PDMS substrate. The white circles show the change in apparent liquid contact angle $\Delta \theta =\theta-\theta_{eq}$, with $\theta_{eq}=105\pm2^\circ$. The grey diamonds show the ridge rotation $\varphi $. The errorbars represent 2 times the standard deviation of the corresponding data. The red line shows the prediction by (\ref{eq:neumann-rotation}). Inset: Same data obtained for wetting on the PVS substrate.}
  \label{fgr:phi_vs_speed_yale}
\end{figure}

We define the low-velocity regime as $v \ll v^*$, where

\begin{equation}\label{eq:vstar}
v^* = \frac{\gamma}{G\tau},
\end{equation}
is the characteristic speed in the problem. This velocity compares the elastocapillary length $\gamma/G$ to the timescale $\tau$ of the gel defined in (\ref{eq:gel1}) (see Appendix~\ref{app:rheology} for details on the rheology of the gel). For the PDMS and PVS substrates, we find $v^*$ of the order of $1~$mm/s. In figure~\ref{fgr:phi_vs_speed_yale} the measurements of the ridge rotation angle $\varphi $ are reported as grey diamonds. On the same figure, we superimpose the change of the liquid angle $\Delta \theta = \theta-\theta_{eq}$. At small velocities $v\ll v^*$, the two measurements perfectly overlap, for both the PDMS substrate (main figure) and the PVS substrate (inset). 

The equality $\Delta \theta = \varphi$ is a central experimental result of the paper. It has two important consequences: (i) the relative angles of dynamical ridges are still given by Neumann's law; (ii) the liquid angle ``passively" follows the rotation of the solid -- implying that the relevant dissipation only takes place inside the solid. These points will be further elaborated in Sec.~\ref{sec:theory}, where we provide a theoretical demonstration that Neumann's law indeed applies as long as the rheological exponent $n<1$.

These observations are in line with the theoretical prediction by~\cite{KarpNcom15}, based on small deformations of the substrate and symmetric solid surface tensions $\Upsilon_s=\Upsilon_{SV}=\Upsilon_{SV}$. For small velocity, the theory simplifies to a power law form~\cite{KarpNcom15},
\begin{equation}\label{eq:smallspeed}
\Delta \theta = \varphi = \frac{2^{n-1}\,n \sin \theta}{\cos\frac{n\pi}{2}}  \, \left(\frac{\gamma}{\Upsilon_s}\right)^{n+1}\, \left(\frac{v}{v^*}\right)^n.
\end{equation}
%
%
In this expression $\gamma$ is the liquid-vapor surface tension, $\Upsilon_s$ is the solid surface tension, $n$ is the  exponent of the gel's loss modulus $G'' \sim \omega^n$. The predicted power law dependence (\ref{eq:smallspeed}), with $n$ and $v^*$ obtained from independent rheological measurements, is in excellent agreement with experimental observations. Importantly, however, the agreement is not fully quantitative. In the model, the solid surface tensions are assumed to be equal, $\Upsilon_s=\Upsilon_{SV}=\Upsilon_{SV}$. Its value is here used as an adjustable parameter, giving $\Upsilon_s=27$mN/m for the PDMS gel ($\Upsilon_s=15$mN/m for PVS). This fitted value is in fact too small to be able to create a Neumann balance with the surface tension of water, which requires that $\Upsilon_s$ is at least $\gamma/2$. A similar lack of quantitative agreement was discussed in experiments of drops sliding down on thin elastic layers~\cite{zhao2018geometrical}. Part of the disagreement can be attributed to the geometrical linearisation of the interface shape in the model, but a fully nonlinear description is still lacking to date -- we will return extensively to this point in sections~\ref{sec:theory} and~\ref{sec:green} of the paper.

\subsubsection{Large velocity.}
\begin{figure}[tb]
\centering
  \includegraphics[width=86mm]{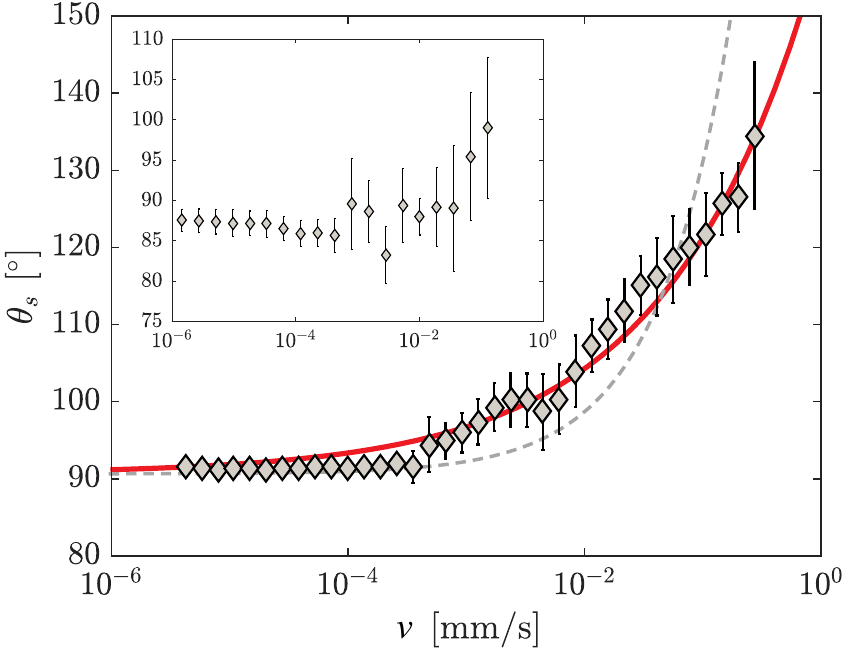} 
  \caption{Solid opening angle $\theta_S$ as function of contact line speed $v$ for water on the PDMS gel. The red and gray-dashed curves shows best fits of power laws $\sim v^{n/2}$ and $\sim v^n$, respectively, where $n\sim 0.58 $ is the exponent of the rheology (see appendix~\ref{app:rheology}).
  Inset: Same data obtained for wetting on the PVS gel.
  }
  \label{fgr:yale_solid_angle}
\end{figure}

Clearly, the experimental data in figure~\ref{fgr:phi_vs_speed_yale} show that a new regime appears when approaching $v^*\sim 1~$mm/s. First, the power-law regime is no longer followed but gives way to a saturation of $\Delta \theta $. More strikingly, however, the angles $\Delta \theta$ and $\varphi$ no longer take on the same value. To interpret these findings, we present the measurement of $\theta_S$ versus instantaneous contact line velocity $v$, shown in figure \ref{fgr:yale_solid_angle}. The main panel shows the result on PDMS substrates, while the inset reports very similar trends observed on PVS. As can be observed, $\theta_S$ is approximately constant at low velocities ($< 10^{-3}$~mm/s). This once more suggests that the solid angle is determined by the surface tensions according to  Neumann's law. At larger velocities, however, $\theta_S$ is found to increase with $v$. This effect was previously reported in \cite{PRLstick-slip}, where the shallower angle was shown to be at the origin of the depinning transition. In contrast to the ridge rotation, which for small speeds follows the same power law as $G''$, so that $\Delta \theta \sim v^n$, the solid angle is much better approximated by the empirical fit $\theta_S\sim v^{n/2}$ (red curve in Fig.~\ref{fgr:yale_solid_angle}).

The intriguing change in $\theta_S$ has important consequences. Firstly, it implies a change of ridge geometry and by consequence, one can no longer unambiguously define a rotation angle $\varphi$. In particular, our choice of using the bisector of the tangent vectors becomes arbitrary, and there is no reason why the determined $\varphi$ would coincide with $\Delta \theta$. Secondly, the change in $\theta_S$ and the validity of Neumann's law (that will be further shown in Sec.~\ref{sec:theory}), points to a dynamical increase of the solid surface tension. Assuming a liquid-vapor surface tension of pure water ($\gamma = 72$~mN/m), we can in fact estimate the solid surface tensions using the Neumann balance, see also~\cite{xu2017direct}.
For the PDMS gel, this gives $\Upsilon_{SV}=42\pm2$~mN/m and $\Upsilon_{SL}=58\pm4$~mN/m at vanishing velocity. The values change to $\Upsilon_{SV}=37\pm4$mN/m and $\Upsilon_{SL}=82\pm7$mN/m for a relatively large velocity ($0.07$ mm/s).

\subsection{Summary and open issues} \label{subsec:open}

From these experimental observations, we draw the following conclusions: 

\begin{itemize}
\item At low velocity, the liquid angle variation $\Delta \theta$ exactly follows the rotation of the ridge $\varphi$ within the error bars of the experiment, providing direct evidence for Neumann's law for moving contact lines. This rejects the hypothesis~\cite{Roche:arxiv19} that viscoelasticity could provide a perfectly localised force at the contact line.
\item At low velocity, both angles exhibit a power law dependence $\Delta \theta=\varphi \sim v^n$, with the exponent $n$ directly given by the scaling of the loss modulus $G'' \sim \omega ^n$.
\item The contact angles are qualitatively described by (\ref{eq:smallspeed}). 
\item This linear theory of Eq.~(\ref{eq:smallspeed}), derived under the restrictive assumptions of small deformations and symmetric surface tensions that are assumed constant, does not quantitatively describe the experiment: the value of the fitted solid surface tension $\Upsilon_s$ is off roughly by a factor 2. We note that the experiment is at large deformations.
\item In addition, at large velocity, the solid angle $\theta_S$ increases significantly with velocity. This points to a variable solid surface tension.
\end{itemize}

Interestingly, a similar quantitative issue with Eq.~(\ref{eq:smallspeed}) was reported for drops sliding down thin viscoelastic layers~\cite{zhao2018geometrical}. As the thickness of the layer was reduced, the linear theory based on ridge rotation leads to an overestimation of the sliding velocity. Zhao~\emph{et al.}\cite{zhao2018geometrical} proposed an alternative approach to the problem, based on a dissipation analysis in the small-velocity regime. Their conclusion, however, is that the liquid angle does not follow the ridge rotation, even at small velocity. This is clearly at odds with the direct experimental observations presented here. In the next paragraphs, we therefore critically re-assess these dissipation arguments; see also~\cite{Karpitschka2018}.

Then let us comment on the velocity-dependent $\theta_S$, and in particular on our interpretation in terms of a velocity-dependent solid surface tension. This phenomenon cannot be attributed to a lack in optical resolution or bulk viscoelastic effects, or a combination thereof. The optical resolution limit is $\sim 60$ times smaller than the static elastocapillary length. In the dynamic case, where the effective modulus $|\mu|$ increases with frequency and thus with velocity, the dynamical elastocapillary length reduces. Nonetheless it remains well above the resolution limit. The characteristic frequency of the bulk response is given by the velocity divided by the distance from the tip. In the range between the optical resolution limit and the elastocapillary length, these frequencies are well resolved in the bulk rheological calibration measurements (see Appendix~A). Deformations on scales below the dynamic elastocapillary length are dominated by surface tension effects. Therefore, it can be excluded that features of the bulk viscoelasticity contribute significantly to the increase in $\theta_S$. The effect of bulk rheology is discussed in detail below, when discussing the boundary condition at the contact angle.

The opening of the ridge angle could, in principle, also be explained by a reduction of the liquid surface tension. A moving contact line of a high energy liquid, like water, is prone to picking up surface active contamination, because these lower the surface free energy of the liquid. Literature reports experiments (though with other types of PDMS substrates), for which the surface tension of water is reduced over time because extractables adsorb to the air-water interface. In the present experiments, however, the same trend of $\theta_S$ was recorded over multiple wetting and dewetting cycles in a single experiment, and after short and long contact times of the liquid with the substrate. This excludes transient effects that stem from a gradual contamination of the air-water interface. Similarly, this excludes poroelastic effects that may occur on long timescales. 

Another important set of recent experiments revealed variations of static contact angles upon stretching the substrate~\cite{xu2017direct,schulman2018surface}. While those experiments were performed at equilibrium, the solid angle $\theta_S$ was found to increase with surface strain, in a way similar to the $\theta_S$ dependence on velocity reported here. Below, we therefore develop a theory that includes the Shuttleworth effect, to explore the influence of variable surface tensions on dynamical wetting ridges. We note here that a surface ``skin'', a thin layer with different mechanical properties than the bulk viscoelastic material, would lead to a strictly equivalent description as long as the layer thickness is much smaller than any other length scale in the problem.

\section{Continuum theory at large deformation} \label{sec:theory}
In this section, we will develop a general formulation for a two-dimensional contact line moving over a viscoelastic substrate. We pose the full mechanical problem based on continuum visco-elasto-capillarity, without relying on the approximations of  small deformation and constant solid surface tension. Particular attention will be paid to the boundary conditions at the contact line.

\subsection{Kinematics and parametrization}

In the standard approach of large deformations, the material is described in a Lagrangian description based on material coordinates $\vec X$, which represents the positions of material points in the reference state.
The deformation is then described by the mapping $\vec x = \vec f(\vec X)$, describing the position of material points in the deformed state.
Capillary effects occur at the free surface.
To describe the nonlinear geometry of the interface, we introduce curvilinear coordinates $S$ and $s$, respectively, for the reference and current states (cf. figure~\ref{fgr:theory}).
Clearly, the surface strain $\epsilon$ is defined by $ds = (1+\epsilon)dS$, while the positions along the free surface $\vec x^*(S)=\vec x(X\!=\!S,Y\!=\!0)$ can be reconstructed as

\begin{equation}
\vec x^*(S) = \vec x^*_0 + \int_{s_0}^s ds \, \vec t = \int_{S_0}^S dS \, (1+\epsilon) \vec t.
\end{equation} 
Here we introduced the tangent vector $\vec t = d\vec x/ds = (\cos \phi,\sin\phi)$, by introducing the interfacial orientation angle $\phi$ that varies along $s$ (or $S$).

In what follows we will make extensive use of the fields $\phi$ and $\epsilon$, which are considered to be functions of $s$.  These fields can be formally defined from derivatives of the map $\vec x = \vec f(\vec X)$ along the interface. The reason for using $\phi$ and $\epsilon$ will become clear below: these are the natural variables to express the surface tractions along the interface of the viscoelastic layer. 

\subsection{Mechanical equilibrium and the Shuttleworth effect}

\begin{figure}[tb]
\centering
  \includegraphics[width=86mm]{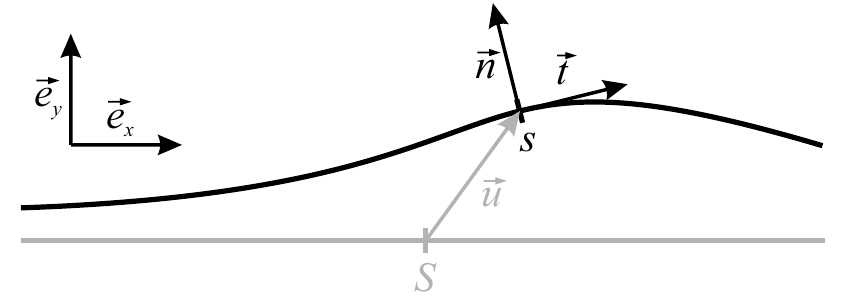}
  \caption{Curvilinear coordinate $s$ along the deformed interface, with the tangential unit vector $\vec t$ and normal unit vector $\vec n$. The curvilinear coordinate in the reference state is denoted $S$, from which we define the displacement field $\vec u$.}
  \label{fgr:theory}
\end{figure}

We consider the dynamics of both the liquid and the solid to be overdamped, such that inertial effects can be neglected. In that case, the Cauchy stress inside the elastic solid ($\boldsymbol{\sigma}$) and inside the liquid ($\mathbf T$) are both divergence-free

\begin{equation}\label{eq:div}
\nabla \cdot \boldsymbol{\sigma} = 0, \quad \nabla \cdot \mathbf T = 0.
\end{equation}
These equations represent the mechanical equilibrium of volume elements, respectively, inside the solid and the liquid. The solid-liquid interaction takes place at the interface, by the boundary condition on the stress. In a two-dimensional description, the interface condition can be written as~\cite{Snoeijer2018}
\begin{equation}\label{eq:tractiondiscontinuity}
\boldsymbol{\sigma}\cdot \vec n - \mathbf T \cdot \vec n = \frac{\partial}{\partial s}\left( \Upsilon_s \, \vec t \right),
\end{equation} 
where we further introduced the surface normal $\vec n= d\vec t/d\phi$ (cf. figure~\ref{fgr:theory}). The right hand side of (\ref{eq:tractiondiscontinuity}) represents the discontinuity of stress due to solid surface tension $\Upsilon_s$. Using the connection $d\vec t/ds = \kappa \, \vec n$, where $\kappa=d\phi/ds$ is the interface curvature, the discontinuity of the normal stress gives the Laplace pressure $\Upsilon_s \kappa$. In the tangential direction, one recognises a Marangoni-like stress $\partial \Upsilon_s/\partial s$ whenever the surface tension is not uniform along the interface. 

Let us elaborate on solid capillarity. In case the interfacial mechanics is non-dissipative\footnote{Dissipative interfacial mechanics can be captured by surface-constitutive relation, similar to bulk viscoelasticity. Then, one needs to take into account also the surface rate of strain, $\dot \epsilon$ or, more generally, a functional dependence on the surface strain history. Such a situation could, for instance, arise in presence of brush-like surface layers that exhibit conformational relaxation in response to  strain.}
, it can be captured by a surface free energy $\gamma_s$. While for simple liquids the surface energy takes on a constant value, this is not necessarily the case for surfaces of (visco)elastic materials, owing to the Shuttleworth effect~\cite{Shuttleworth1950a, Andreotti2016a,style2017elastocapillarity}. Namely, the surface energy $\gamma_s(\epsilon)$ is in general expected to be a function of the surface strain $\epsilon$, a fact that was verified explicitly for polymeric interfaces~\cite{xu2017direct,schulman2018surface,Snoeijer2018}. Then, surface tension $\Upsilon_s$ is not equal to the surface energy $\gamma_s$, but follows from the Shuttleworth equation~\cite{Shuttleworth1950a}, 
\begin{equation}\label{eq:Shuttleworth}
\Upsilon_s = \gamma_s + (1+\epsilon) \frac{d\gamma_s}{d\epsilon}.
\end{equation} 
In general, one thus cannot assume a priori that surface tension takes on a constant value.
In addition, the Shuttleworth effect implies that the Marangoni stress in (\ref{eq:tractiondiscontinuity}) needs to be taken into account.

We now work out the case where the liquid traction $\mathbf T=0$, i.e. Laplace pressure and viscous stress inside the drop can be neglected, and where $\Upsilon_s$ depends instantaneously on $\epsilon$, i.e. non-dissipative surface mechanics. Writing (\ref{eq:tractiondiscontinuity}) in normal and tangent directions, we find
\begin{equation}\label{eq:symmetry}
 \vec \sigma = \epsilon' \frac{d\Upsilon_s}{d\epsilon} \, \vec t + \phi' \Upsilon_s \, \vec n.
\end{equation}
Here we introduced the surface traction vector $\vec \sigma \equiv \boldsymbol{\sigma} \cdot \vec n$, while the prime indicates derivative with respect to $s$. Hence, (\ref{eq:symmetry}) nicely shows that the variables $\epsilon'$ and $\phi'$ appear symmetrically: they respectively provide the forcing tangential and normal to the interface. 

\subsection{The wetting boundary conditions}\label{sec:bc}

Equation~(\ref{eq:symmetry}) contains first order derivatives of $\epsilon$ and $\phi$ with respect to $s$. Thus, on singular points with discontinuities in $\epsilon$ or $\phi$, it needs to be complemented by corresponding boundary conditions.
In the wetting problem, such discontinuities indeed arise at the contact line. 
The wetting boundary conditions on $\phi$ and $\epsilon$ are obtained from the degrees of freedom that define the position of the three phase contact line.
We consider the contact line to be at the location $\vec x_{\rm cl}$, which corresponds to the material coordinate $S_{\rm cl}=R$. The two degrees of freedom are thus the Eulerian position $\vec x_{\rm cl}$ and the Lagrangian position $R$. 

Since $\vec x_{\rm cl}$ is associated to a spatial coordinate, its displacement will involve a (vector) force balance -- this will be the Neumann balance, which serves as a discontinuity in the contact angles $\phi^\pm=\phi(R^\pm)$. By contrast, a change of the degree of freedom $R$ will involve the exchange of material across the contact line. This will give rise to a so-called ``configurational balance", which, as it involves material exchange, we will refer to this as a balance of chemical potential. This provides a boundary condition on $\epsilon^\pm$.

\subsubsection{Boundary condition on $\phi$: When is Neumann's law valid?}\label{subsec:neumann}~
From a theoretical point of view, we can now proceed along two distinct routes, that lead to the same result. In the first route, we represent the capillary action of the liquid-vapor interface as a highly localised traction, pulling with a tension $\gamma$ along the direction $\vec t_{LV}$ of the liquid-vapor interface. Treating this liquid traction as a perfectly localised Dirac $\delta$-function at $s=s_{\rm cl}$, 

\begin{equation}
\mathbf T \cdot \vec n = \gamma  \vec t_{LV}\;\delta \left( s - s_{\rm cl}\right),
\end{equation}
equation (\ref{eq:tractiondiscontinuity}) can be written as~\cite{Limat2012a,Lub14,Style2012a}
\begin{equation}\label{eq:delta}
\vec \sigma = \frac{\partial }{\partial s}\left( \Upsilon_s \, \vec t \right)+  \gamma  \vec t_{LV}\;\delta \left( s - s_{\rm cl}\right).
\end{equation}
This provides a way to derive Neumann's law for the contact angles, by integrating over an arbitrarily small distance across the contact line, from $s_{\rm cl}^-$ to $s_{\rm cl}^+$. This gives
\begin{align}\label{eq:deltaint}
\int_{s_{\rm cl}^-}^{s_{\rm cl}^+} ds \, \vec \sigma &= \gamma \vec t_{LV} + \left( \Upsilon_s \, \vec t \right)^+ - \left( \Upsilon_s \, \vec t \right)^- \nonumber \\
&= \gamma \vec t_{LV} +  \Upsilon_{SV} \vec t_{SV}+ \Upsilon_{SL} \vec t_{SL},
\end{align}
where in the last line we expressed the values on either sides of the contact line by using the indices for the solid-vapor (SV) and solid-liquid (SL) interfaces.

The second, more formal route to (\ref{eq:deltaint}) is based on variational principles~\cite{Snoeijer2018}. In that case, one does not represent the liquid-vapor interface by a $\delta$-shaped traction. Instead, one explores the work done by a virtual displacement of the Eulerian contact line position. Application of the virtual work principle then leads to (\ref{eq:deltaint}) as a boundary condition that needs to be imposed at the contact line~\cite{Snoeijer2018}. 

We can now infer an important conclusion: in the boundary condition (\ref{eq:deltaint}), the bulk viscoelastic traction is integrated over an infinitesimally small distance that crosses the contact line. As long as the traction exhibits a divergence that is weaker than $1/(s-s_{\rm cl})$, the integrated contribution $\int ds \, \vec \sigma $ vanishes in the limit of an infinitesimal integration domain, in which case it can be omitted. So, depending on the degree of singularity of $\vec \sigma$, the boundary condition (\ref{eq:deltaint}) can indeed reduce to a vectorial balance of the three surface tensions, which is the Neumann law.

As an interesting example of a non-integrable traction, we quote the case of Newtonian liquids that spread on rigid surfaces. In that case, the viscous stress $\boldsymbol{\sigma} \sim \eta v/(s-s_{\rm cl})$, where $\eta$ is the liquid viscosity and $v$ the contact line speed. This non-integrable stress gives rise to the famous ``moving contact line singularity"~\cite{Huh1971a,RevBonn09,RevSnoB13}, which is also encountered for motion over a Kelvin-Voigt viscoelastic solid~\cite{KarpNcom15}. 

By contrast, for a purely elastic substrate the formation of a sharp corner at its surface with a solid angle $\theta_S$ leads to a much weaker, logarithmic singularity of  $\vec \sigma$; both in linear elasticity~\cite{Johnson,Lub14} and at large deformation~\cite{Singh1965}. This renders the elastic stress integrable, in the sense that its contribution in (\ref{eq:deltaint}) vanishes when taking the limit of infinitesimal integration domain. This implies the validity of Neumann's law for elastic media. 

For viscoelastic media, one can show that the stress remains integrable when the rheological exponent $n<1$. Namely, at a typical distance $\ell$ from the contact line, the material is excited at a frequency $v/\ell$ and the associated stress $\sigma \sim \mu (v\tau/\ell)^n$. The integrated stress contribution indeed vanishes when $n<1$, so that Neumann is still valid. The validity of this scaling argument will be shown explicitly in Appendix~\ref{app:linear}, for the linear theory~\cite{KarpNcom15}, as well as for the dissipation theory~\cite{Karpitschka2018}. Since for all experimental substrates used in the literature the rheological exponent $n<1$, we conclude that Neumann's law will hold even dynamically. This is in agreement with the experimental observation $\Delta \theta = \varphi$.

\subsubsection{Boundary condition on $\epsilon$.}~The boundary condition on the surface strain originates from exchange of material across the contact line, induced by a change of the Lagrangian contact line position $R$. For homogeneous elastic media without any pinning sites, such a change of contact line position should be neutral with respect to the total energy of the system. For the purely elastic case, it has been shown from variational principles that this neutrality provides a boundary condition~\cite{Snoeijer2018}, 
\begin{equation}\label{eq:mubis}
\mu_{SV}=\mu_{SL}, \quad  \quad {\rm with} \quad 
\mu = (1+\epsilon)^2 \gamma'(\epsilon) + \mu_{\rm bulk} 
\end{equation}
This is to be interpreted as an equality of chemical potential, $\mu^+=\mu^-$, describing the exchange of material between two domains. In the following we will assume a local thermodynamic equilibrium with respect to material exchange, so that we can employ the same condition in the dynamical case.

The first term in (\ref{eq:mubis}) is a surface term, involving $\gamma'$ and is therefore directly related to the Shuttleworth effect. The second term, $\mu_{\rm bulk}$, is associated to changes in elastic energy associated to the change in configuration below the contact line.  Such a term would play a role when substrate inhomogeneities are present, such as pinning sites, dislocations or other microscopic effects beyond the continuum description. However, for materials that are homogeneous in the reference state, i.e. prior to the application of a droplet, the continuum framework must give $\mu_{\rm bulk}=0$. This can be inferred from Eshelby's considerations~\cite{Eshelby1975}, who applied Noether's theorem to the translational invariance in the ``space of material coordinates". Homogeneous media are translationally invariant in material space, and thus cannot lead to a change in energy upon a change in conformation -- irrespective of the deformation. Hence, with $\mu_{\rm bulk}=0$ we find the boundary condition
\begin{equation}\label{eq:mu}
(1+\epsilon^+)^2 \gamma'(\epsilon^+) = (1+\epsilon^-)^2 \gamma'(\epsilon^-),
\end{equation}
to be imposed at the contact line. This shows that the presence of a Shuttleworth effect can lead to discontinuities of strain.

To illustrate the implications of (\ref{eq:mu}), let us expand the solid surface energies for small strain up to quadratic order, 
\begin{align}\label{eq:constitutive1}
\gamma_{SV}&=\gamma_{SV}^0+\gamma_{SV}^1\epsilon+\frac12 \gamma_{SV}^2\epsilon^2\\
\gamma_{SL}&=\gamma_{SL}^0+\gamma_{SL}^1\epsilon+\frac12 \gamma_{SL}^2\epsilon^2.\label{eq:constitutive2}
\end{align}
Then, the Shuttleworth equation (\ref{eq:Shuttleworth}) provides the surface stress:
\begin{align}
\Upsilon_{SV} &= (\gamma_{SV}^0+\gamma_{SV}^1)+(2\gamma_{SV}^1+\gamma_{SV}^2)\epsilon+\frac32 \gamma_{SV}^2\epsilon^2\\
\Upsilon_{SL} &= (\gamma_{SL}^0+\gamma_{SL}^1)+(2\gamma_{SL}^1+\gamma_{SL}^2)\epsilon+\frac32 \gamma_{SL}^2\epsilon^2.
\end{align}
The boundary condition (\ref{eq:mu}) then becomes,
\begin{equation}
(1+\epsilon_{SL})^2\left(\gamma_{SL}^1+ \gamma_{SL}^2 \epsilon_{SL}\right)=(1+\epsilon_{SV})^2\left(\gamma_{SV}^1+ \gamma_{SV}^2 \epsilon_{SV}\right)
\end{equation}
which should be interpreted as a condition that relates $\epsilon_{SL}$ and $\epsilon_{SV}$ on either sides of the contact line. 

It is now apparent that whenever the coefficients $\gamma_{SV}^1 \neq \gamma_{SL}^1$, this will lead to a Shuttleworth-induced strain-discontinuity across the contact line,  i.e. $\epsilon_{SV} \neq \epsilon_{SL}$. This effect is in direct analogy to the discontinuity of density across a liquid-vapor interface, that is governed by the equality of chemical potential (associated to material exchange). Without the Shuttleworth effect, the boundary condition (\ref{eq:mu}) is automatically satisfied and $\epsilon$ will be continuous across the contact line. By constrast, a large difference in $\gamma^1$ implies values of $\epsilon$ to be of order unity, and thus beyond the expansion of~(\ref{eq:constitutive1}),~(\ref{eq:constitutive2}).

\section{Green's function approach}\label{sec:green}

Up to now the developments have been exact, and account for the geometric nonlinearities associated to large deformations. To make concrete predictions, previous theoretical approaches have considered a viscoelastic Green's function description~\cite{LALLang96,Karpitschka2018,zhao2018geometrical}, where the interface deformations of the substrate are computed from tractions provided at the boundary. In what follows we will extend the Green's function formalism, i.e. still using linear viscoelastic response, while retaining the geometric nonlinearities associated to capillarity -- including the Shuttleworth effect.

\subsection{Dynamic Green functions}
For an initially flat substrate, the problem naturally separates into horizontal and vertical directions, respectively along $\vec e_x$ and $\vec e_y$. The common Green's function approach establishes a linear relation between the tractions $\vec \sigma$ and the displacements $\vec u$ at the free surface. Subsequently, the linear response of the viscoelastic layer can, in its most general form, be written as
\begin{align}\label{eq:greensystem1}
\vec \sigma \cdot \vec e_x &= K_{x\epsilon} \otimes \epsilon + K_{x\phi} \otimes \phi \\
\vec \sigma \cdot \vec e_y &= K_{y\epsilon} \otimes \epsilon + K_{y\phi} \otimes \phi. \label{eq:greensystem2}
\end{align}
In these expressions $\otimes$ indicates a convolution over the entire free surface, and over the entire history of deformation for viscoelastic substrates. The $K_{\alpha\beta}$ are the associated dynamical Green's functions, which can be obtained by rewriting, in the small-deformation limit, the strain as a function of the displacement $\vec u$, $\epsilon = \partial u_x/\partial x$ and $\phi = \partial u_y/\partial x$.

The rigorous validity of the linear response of the surface traction to $\phi$ and $\epsilon$ is valid only for small deformations. However, here we use the Green's function approach as a particular constitutive relation, which enables us for the first time to highlight the geometric nonlinearities associated to solid capillarity, including the Shuttleworth effect. These geometric nonlinearities in $\phi$ and $\epsilon$ show up in the forcing on the right of (\ref{eq:symmetry}), due to the misalignment between $\vec n$ and $\vec e_y$, and between $\vec t$ and $\vec e_x$. At the contact line, boundary conditions on $\epsilon$ and $\theta$ must be employed to close the problem.

\begin{figure*}[t!]
\centering
\includegraphics{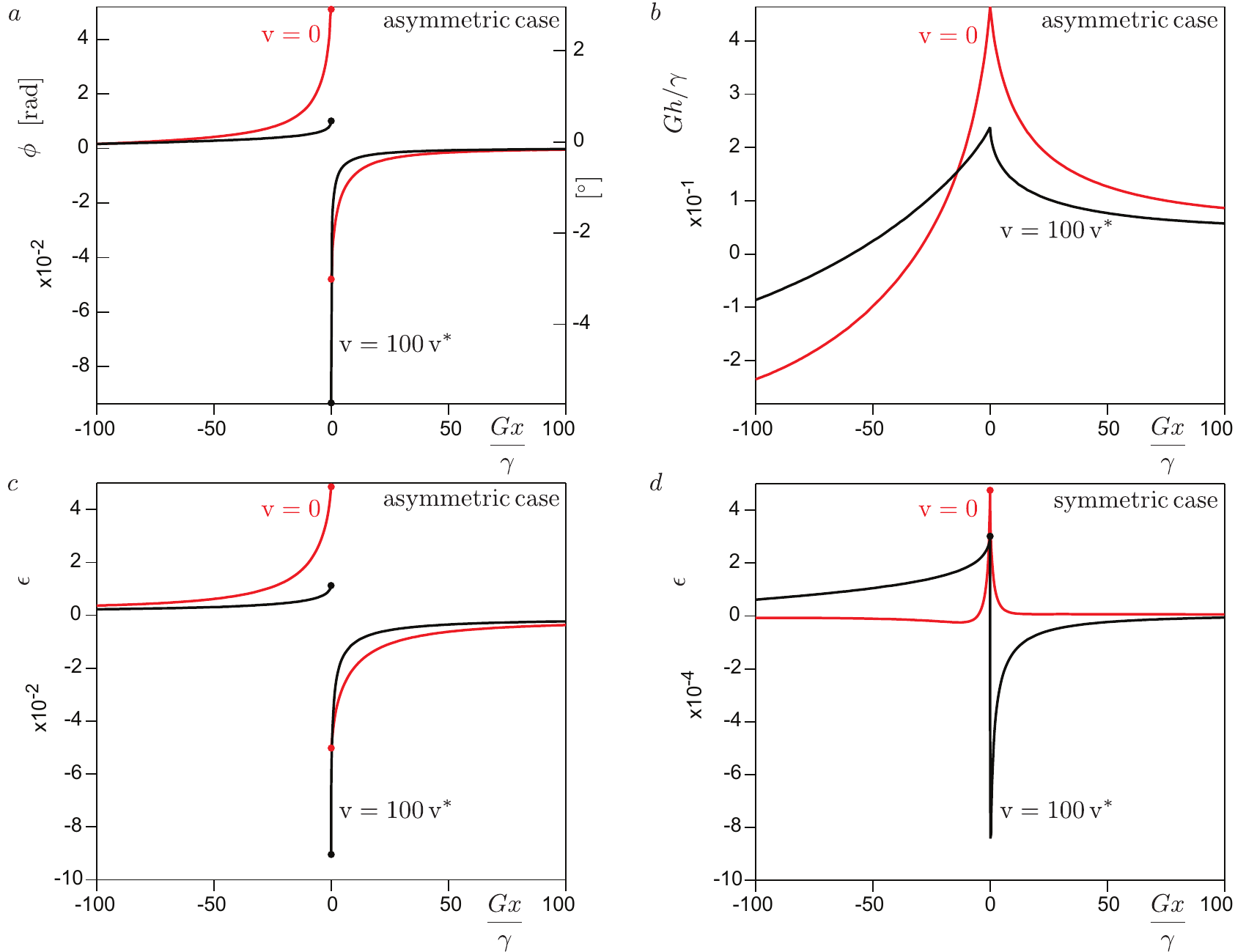}
 \caption{Typical static and dynamic solutions with Shuttleworth effect, on an infinitely thick substrate. Red curves are $v=0$ and black curves are $v=100v^*=100 \gamma/(G \tau)$. For panels (a-c),  $\gamma_{SV}^0=5.5\,\gamma$, $\gamma_{SL}^0=4.5\,\gamma$, $\gamma_{SV}^1=5.5\,\gamma$, $\gamma_{SL}^1=4.5\,\gamma$, $\gamma_{SV}^2=\gamma_{SL}^2=10\,\gamma$. (a) Profiles of the free surface angle $\phi$ as a function of $x$. (b) Corresponding ridge shapes $h(x)$ obtained by integration of $\tan \theta$ over $x$. (c) Surface strain $\epsilon(x)$ for the same conditions. (d) Same, but for symmetric surface conditions conditions $\gamma_{SV}^0=\gamma_{SL}^0=5\,\gamma$, $\gamma_{SV}^1=\gamma_{SL}^1=5\,\gamma$, $\gamma_{SV}^2=\gamma_{SL}^2=10\,\gamma$.}
  \label{fgr:TypicalSolution}
\end{figure*}

\subsection{Travelling wave solutions}
We now describe the technical aspects of treating the system (\ref{eq:greensystem1},\ref{eq:greensystem2}). A first step is to take spatial Fourier transforms of these expressions, so that the convolution reduces to a product. For the dynamics of viscoelastic layers, governed by a dynamic modulus $\mu(\omega)=G'(\omega)+iG''(\omega)$, the time-evolution is dealt with in similar fashion, namely by taking a temporal Fourier transform~\cite{LAL96,LALLang96,KarpNcom15}. Indicating the transform in space by ``tilde", with variable $q$, and in time by ``hat", with variable $\omega$, then the Green's functions are of the form:
\begin{equation}
\hat{\tilde{K}}_{\alpha\beta}(q,\omega) = \frac{\mu(\omega)}{k_{\alpha \beta}(q)},
\end{equation}
where $k_{\alpha \beta}(q)$ and $\mu(\omega)$ are the spatial green's function representing the substrate geometry, and the complex shear modulus, respectively. When considering contact lines that propagate at a constant velocity, we can further simplify these expressions using travelling wave solutions

\begin{align}%
\vec \sigma(x,t) &= \vec \sigma_c(\bar x) \\
\epsilon(x,t) &= \epsilon_c(\bar x) \\
\phi(x,t) &= \phi_c(\bar x),\\
\mathrm{with}\quad \bar x &= x - vt.
\end{align}%
where $v$ is the wave velocity.
The corresponding spatio-temporal Fourier transforms of the deformations read
\begin{align}
\hat{\tilde{\epsilon}}(q,\omega) &= \tilde{\epsilon}_c(q) \, 2\pi \delta(\omega-qv) \\
\hat{\tilde{\phi}}(q,\omega) &= \tilde{\phi}_c(q) \, 2\pi \delta(\omega-qv).
\end{align}
The appearance of the $\delta$-functions is convenient. Namely, after applying the Green's functions, the resulting $\hat{\tilde{\sigma}}(q,\omega)$ can be transformed back to the temporal domain using $\omega=qv$. Hence, we finally have
\begin{align}\label{eq:greensystem3}
\tilde{\vec \sigma}_c \cdot \vec e_x &= 
\mu(qv) \left[ k^{-1}_{x\epsilon}(q)  \, \tilde \epsilon_c(q) 
+ k^{-1}_{x\phi}(q)  \, \tilde \phi_c(q)\right]\\
\tilde{\vec \sigma}_c \cdot \vec e_y &= 
\mu(qv) \left[ k^{-1}_{y\epsilon}(q)  \, \tilde \epsilon_c(q) 
+ k^{-1}_{y\phi}(q)  \, \tilde \phi_c(q)\right].
\label{eq:greensystem4}
\end{align}
This is a purely spatial form, where the dynamics is included through the parameter $v$ that couples to the substrate's rheology.

To numerically solve for the dynamic shape of the gel, we consider the infinite thickness limit of incompressible media. In this case the $k_{\alpha \beta}$ are diagonal such that the constitutive equations reduce to~\cite{Johnson}:
\begin{align}
\tilde{\vec \sigma}_c \cdot \vec e_x &=-2i \mathcal{S}(q) \mu(qv) \tilde{\epsilon}_c(q)\\
\tilde{\vec \sigma}_c \cdot \vec e_y &=-2i \mathcal{S}(q) \mu(qv) \tilde{\phi}_c(q),
\end{align}
where $\mathcal{S}(q)$ is the sign of $q$. We choose $\mu(\omega)$ as defined in (\ref{eq:gel1}). In Appendix~\ref{app:linear} we for completeness show how the asymptotic result Eq.~(\ref{eq:smallspeed}) can be derived in the limit of small $\phi,\epsilon$ with equal surface tensions without Shuttleworth effect. For the more general numerical solutions presented here, we consider a two-dimensional drop deposited on the gel, which therefore presents one advancing and one receding contact line. In (\ref{eq:tractiondiscontinuity}) one has to introduce the liquid traction $\mathbf T \cdot \vec n$ associated with a constant Laplace pressure inside the drop, acting purely along $\mathbf n$. Writing this traction as $\partial \vec P/\partial s$, we write (\ref{eq:tractiondiscontinuity}) as
\begin{equation}\label{eq:numericalstuff}
\vec \sigma = \cos \phi \frac{d }{d x}\left( \Upsilon_s \, \vec t +\vec P \right).
\end{equation}
It is well known~\cite{Style2012a,Lub14} that the pressure inside the drop has no effect on the contact line region if the drop size is large compared to the elastocapillary length $\gamma/G$, which is the case here. In order to simplify the calculation and to optimise the code, we therefore take $\vec P$ as a linear function of $x$ inside the drop (as would be the case on a rigid substrate), while it vanishes outside the drop. Then, the strength of the jump in $\vec P$ in (\ref{eq:numericalstuff}) gives rise to a discontinuity. By adjusting the jump in $\vec P$ at the contact line, one can impose the desired discontinuities in $\phi$ and $\epsilon$ as dictated by the boundary conditions. As the equations are non-linear, we iteratively solve them by computing the difference between the imposed traction and the elastic stress given by the Green's function.

\subsection{Numerical solutions with Shuttleworth effect}

We now present numerically obtained dynamic wetting ridges, in the presence of the Shuttleworth effect. The surface properties are defined by (\ref{eq:constitutive1},\ref{eq:constitutive2}), which involves the coefficients $\gamma^0,\gamma^1,\gamma^2$ for both the solid-liquid and solid-vapor interfaces. Rather than fitting the experimental data with these 6 parameters, we highlight the possible scenarios via various illustrations for representative choices of the surface energy.

Figure~\ref{fgr:TypicalSolution} shows a typical solution, with parameters chosen to give a large discontinuity of the surface strain $\epsilon$ at the contact line (numerical parameters quoted in the caption). The red data correspond to the static solution with $v=0$, while the black data are for case of very large velocity $v=100 v^*$, where we remind $v^*=\gamma/(G\tau)$. The profiles for the local surface angle $\phi(x)$ and surface strain $\epsilon(x)$ are respectively given in figures~\ref{fgr:TypicalSolution}(a,c), and for this particular choice of parameters, these exhibit very similar behaviours. The boundary conditions enforce a discontinuity for both $\phi$ and $\epsilon$, and both exhibit a sharp peak on both sides of the contact line. Once integrated, one recovers the shape $h(x)$ of the interface, given in figure~\ref{fgr:TypicalSolution}(b). The evolution from static to dynamic profile strongly resembles that observed in experiment (figure~\ref{fgr:ridge_shapes}). When the velocity is increased, the wetting ridge rotates, similar to the experimental observation. For comparison, in figure~\ref{fgr:TypicalSolution}(d) we also provide a ``symmetric" example with $\gamma_{SV}=\gamma_{SL}$. In that case, $\epsilon$ is continuous across the contact line and the typical values for surface strain remain very small. 

\begin{figure}[tb!]
\centering
  \includegraphics{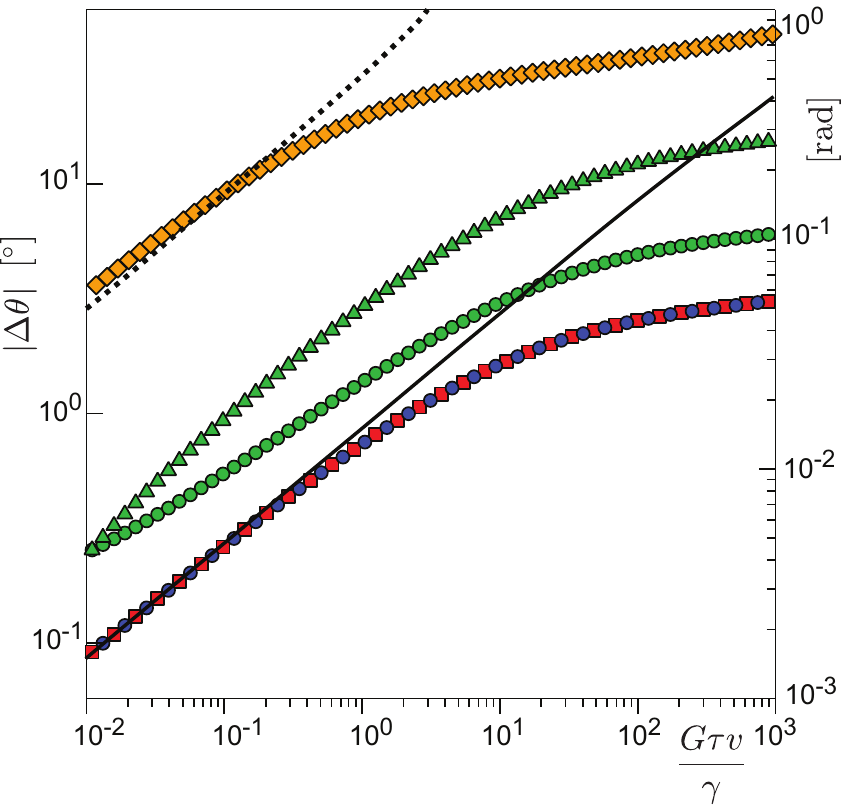}
  \caption{Dependence of the liquid angle $\Delta \theta= \theta - \theta_{eq}$ on the velocity $v$ rescaled by $v^*=\gamma/G\tau$. Left vertical axis indicates degrees, right vertical axis radians. Red squares: without any Shuttleworth effect, for $\Upsilon_{SV}=\Upsilon_{SL}=10 \gamma$. The solid line is the analytical result~(\ref{eq:smallspeed}) at small velocity and small strain. Orange diamonds: without any Shuttleworth effect, for $\Upsilon_{SV}=\Upsilon_{SL}= 1 \gamma$. The dotted line is the power law~(\ref{eq:smallspeed}) for small velocity and small strain, slightly below the numerical data due to geometric nonlinearity. Blue circles: with Shuttleworth effect and symmetric surface energies, $\gamma_{SV}^0=\gamma_{SL}^0=5\,\gamma$, $\gamma_{SV}^1=\gamma_{SL}^1=5\,\gamma$, $\gamma_{SV}^2=\gamma_{SL}^2=10\,\gamma$. The symbols are obtained in both advancing and receding direction. Green circles and triangles:  with Shuttleworth effect and asymmetric surface energies,  $\gamma_{SV}^0=5.5\,\gamma$, $\gamma_{SL}^0=4.5\,\gamma$, $\gamma_{SV}^1=5.5\,\gamma$, $\gamma_{SL}^1=4.5\,\gamma$, $\gamma_{SV}^2=\gamma_{SL}^2=10\,\gamma$.  Triangles: advancing ($v>0$). Circles: receding ($v<0$).}
  \label{fgr:ThetaVsVelocity_Sim}
\end{figure}

Figure~\ref{fgr:ThetaVsVelocity_Sim} shows the resulting dependence of the liquid angle $\Delta \theta = \theta- \theta_{eq}$ on velocity, determined numerically in different situations. As a validation of the numerical technique, we first consider a case without any Shuttleworth effect and symmetric surface tensions, in the regime of small deformation (achieved for $\gamma/\Upsilon_s\ll 1$). The result is  represented by the red squares in figure~\ref{fgr:ThetaVsVelocity_Sim}, which is in perfect agreement with the small speed asymptotics (\ref{eq:smallspeed}), superimposed as the solid line. A second test is provided by the blue circles, which corresponds to a situation with a Shuttleworth effect, but where $\gamma_{SV}$ and $\gamma_{SL}$ have exactly the same dependence on $\epsilon$. As the strain is negligible in this situation (since no discontinuity of $\epsilon$ will emerge), the blue and red data coincide within error bars. As a third case, the orange diamonds are obtained without any Shuttleworth effect, but with angles of order $1$. This time, the analytical solution (\ref{eq:smallspeed}) is significantly off the numerical points -- at a given value of $\Delta \theta$, the linear theory overestimates the sliding velocity in a way similar to the experiments in~\cite{zhao2018geometrical}. This clearly shows that one cannot expect a fully quantitative agreement between linear theory and experimental data obtained with large deformations of the interface. 

\begin{figure}[tb!]
\centering
  \includegraphics{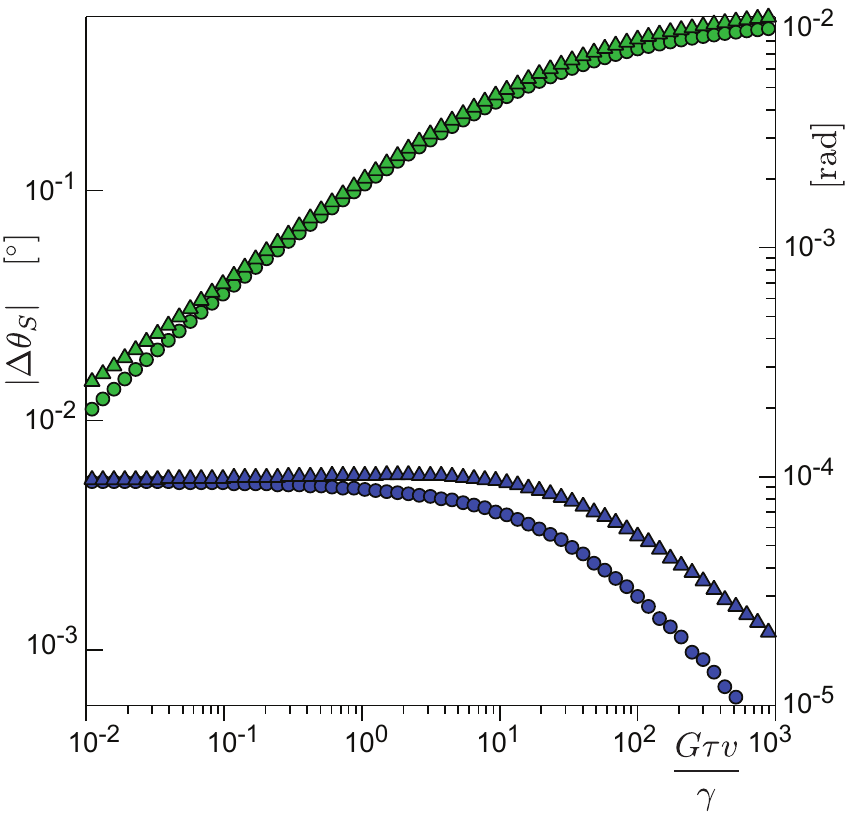}
  \caption{Dependence of the solid angle $\Delta \theta_S= \theta_S-\theta_{S,v=0}$, on the velocity $v$ rescaled by $v^*=\gamma/G\tau$. Left vertical axis indicates degrees, right vertical axis radians. Triangles: advancing ($v>0$). Circles: receding ($v<0$). Blue symbols: with Shuttleworth effect and symmetric surface energies, $\gamma_{SV}^0=\gamma_{SL}^0=5\,\gamma$, $\gamma_{SV}^1=\gamma_{SL}^1=5\,\gamma$, $\gamma_{SV}^2=\gamma_{SL}^2=10\,\gamma$. The symbols are data obtained in both advancing and receding direction. Green symbols:  with Shuttleworth effect and asymmetric surface energies,  $\gamma_{SV}^0=5.5\,\gamma$, $\gamma_{SL}^0=4.5\,\gamma$, $\gamma_{SV}^1=5.5\,\gamma$, $\gamma_{SL}^1=4.5\,\gamma$, $\gamma_{SV}^2=\gamma_{SL}^2=10\,\gamma$.}
  \label{fgr:ThetaSvsVelocity}
\end{figure}

Beyond these tests of self-consistency, the green symbols in the figure~\ref{fgr:ThetaVsVelocity_Sim} represent the most general case, with an asymmetric Shuttleworth effect. Due to the large values of $\epsilon$ on both sides of the contact line, the advancing and receding curves present an asymetry. They both exhibit a regime at small velocity where the rotation angle follows a power law with velocity, and a saturation at large $v/v^*$. However, the main consequence of an asymmetric Shuttleworth effect between both sides of the contact line is the dependence of the solid angle $\theta_S$ with velocity, as illustrated by the green data in Fig.~\ref{fgr:ThetaSvsVelocity}. $\theta_S$ behaves in a similar way as the variation of the liquid contact angle: it presents a power law scaling as $v^n$ at asymptotically small velocity and saturates at large $v/v^*$. This effect is indeed observed experimentally, but with a significantly smaller apparent exponent that remains to be explained. By contrast, when the Shuttleworth effect is symmetric, $\epsilon$ remains small so that the variation of $\theta_S$ with velocity are negligible as well (blue symbols).

\section{Conclusion}\label{sec:discussion}

In summary, we presented detailed experiments on dynamical wetting ridges, quantifying both the liquid and solid contact angles during spreading on two different substrates. These experiments are complemented by theoretical developments that offer a systematic introduction of large deformations and the Shuttleworth effect into the theory of dynamical wetting. 

The results presented here provide a clear answer to the question posed in the title: Namely, Neumann's law is still valid for dynamical wetting on viscoelastic substrates. This follows from the direct measurements of the ridge rotation as a function of contact line speed, which perfectly follows the change of the liquid angle. In the theoretical development, we have shown analytically and  numerically that Neumann's law prevails due to the weak singularity of viscoelastic stress -- provided that the rheological exponent $n<1$. Consistently, this condition on the rheology appears both in the mechanical framework as well as from a dissipation analysis. In principle, viscoelasticity could alter Neumann's law, as was recently hypothesized~\cite{zhao2018geometrical,Roche:arxiv19} -- however, this hypothesis can be rejected for the soft polymeric networks used experimentally, which typically have exponents $n \approx 0.5$.

Several open issues remain. At larger velocities, it is observed that the solid angle $\theta_S$ increases with respect to its equilibrium value. This implies that the solid surface tension is not constant, but depends on velocity. We have investigated whether the Shuttleworth effect, i.e. a strain-dependent surface energy, could explain these observations. While our numerics qualitatively reproduced the increase of $\theta_S$, the observed scaling with velocity is not captured. We hypothesise that the surface tension could depend on strain-rate, which would necessitate a more detailed description in terms of surface rheology. 

For a fully quantitative description, it will be important to develop a nonlinear description of the substrate, beyond the Green's function numerical results presented here. However, the general formulation that we developed clearly highlights the symmetric roles of the interface angle $\phi$ and the surface strain $\epsilon$, both of which are expected to represent a discontinuity at the contact line. It is therefore important that future experiments not only report the shape of the wetting ridge, but also accurately capture the strain along the surface.

\emph{Acknowledgments.~---~}
We are grateful to Anupam Pandey, Harald van Brummelen and Rob Style for discussions on the singularity near the contact line. This work was financially supported by the ANR grant Smart and ERC (the European Research Council) Consolidator Grant No. 616918.

\appendix

\section{Gel rheology}\label{app:rheology}
The gel rheology is measured using an Anton Paar MCR 502 rheometer, in a parallel plate geometry, using a frequency sweep at 1\% strain (both elastomers show a linear response up to 100\% strain). The gel, prepared by mixing the prepolymers, is cured inside the rheometer using the same batch and following the same procedure as the gel used for the experiments. The PVS gel is cured at room temperature for approximately 10 hours. The PDMS requires an additional step to be fully reticulated: the temperature is increased to $80^\circ C$ for 200 minutes.

\begin{figure}
\centering
  \includegraphics[width=86mm]{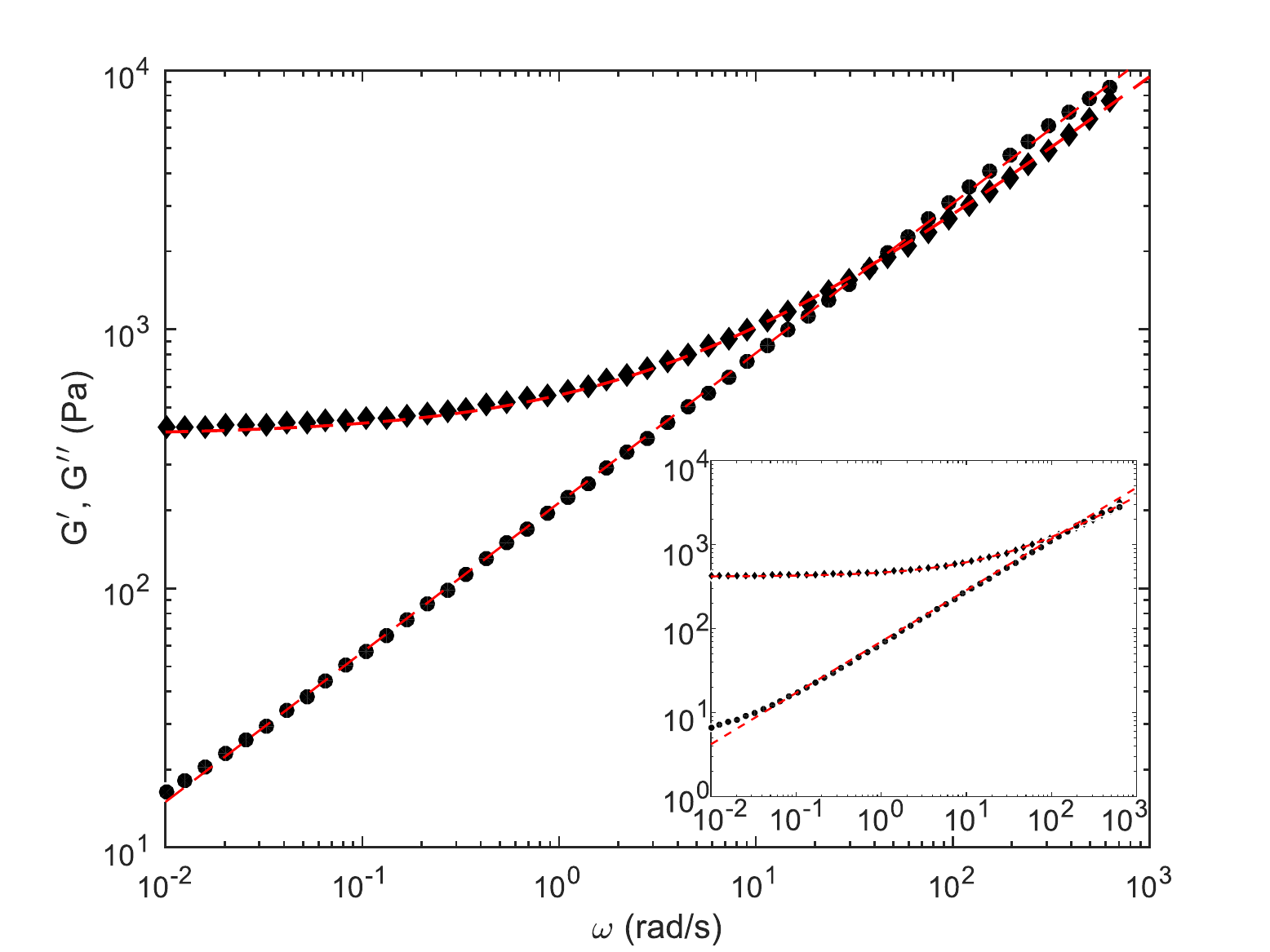}
  \caption{Measured PDMS gel rheology. The diamonds show the storage modulus, the circles show the loss modulus. The fit for equation \ref{eq:gel} is shown in the lines resulting in: $n=0.58$, $G=390 Pa$, $\tau=0.54s$. The inset shows the measured rheology for the PVS gel, with the following fit parameters: $n=0.61$, $G=415 Pa$, $\tau=0.08s$}
  \label{fgr:rheology}
\end{figure}

The measured rheology for both gels are shown in figure \ref{fgr:rheology}, where we can see that both gels behave similarly. The storage modulus, $G'(\omega)$, approaches a finite value at vanishing frequency $\omega$, which defines the shear modulus according to $G\equiv G'(0)$. This means that the gels are elastic in the long time limit, and hence can be classified as viscoelastic solids. The loss modulus, $G''(\omega)$, exhibit a power-law behaviour over nearly the entire frequency range. The rheology of the gels can in fact be accurately fitted by a complex modulus

\begin{eqnarray}\label{eq:gel}
\mu(\omega) = G' + iG'' = G\left[1+(i\omega \tau)^n\right],
\end{eqnarray}
which in terms of storage and loss modulus reads

\begin{align}\label{eq:gelbis}
G'(\omega) &= G \left[ 1 + (\omega \tau)^n \cos\left(\frac{n\pi}{2}\right)\right]\nonumber  \\
G''(\omega) &= G  \sin\left(\frac{n\pi}{2}\right) (\omega \tau)^n.
\end{align}
These satisfy the Kramers-Kronig relations, since the complex modules derives from a stress relaxation function

\begin{equation}
\Psi(t) = G\left[ 1 + \Gamma(1-n)^{-1} \left( \frac{\tau}{t}\right)^n\right].
\end{equation}
Here, $G$ is the static shear modulus, $\tau$ is the response time of the gel and $\omega$ is the excitation frequency. These constants were determined by fitting equation (\ref{eq:gel}) to the measured rheology. The fit was done by finding $n$ from the loss modulus $G''$, which for (\ref{eq:gelbis}) is a pure power-law, and using a least squares fit to find $G$ and $\tau$ from $G'$. The PDMS has the following properties: $n=0.58$, $G=390 Pa$, $\tau=0.54s$, while for the PVS gel we find $n=0.61$, $G=415 Pa$, $\tau=0.08s$.

\section{Analytical solution without Shuttleworth effect}\label{app:linear}

For completeness we recall the result of~\cite{KarpNcom15}, showing how (\ref{eq:smallspeed}) emerges in the present framework in the limit of small deformations, equal surface tensions and no Shuttleworth effect. This is complemented by an explicit calculation of $\Delta \theta$ from a dissipation approach showing that it yields the same result, provided that the rheological exponent $n<1$. Hence, it constitutes a formal proof of Neumann's law for viscoelastic media, provided that $n<1$.

\subsection{Ridge shape}
In the absence of a Shuttleworth effect and for small deformation, the normal and tangential forcing decouple to leading order. Hence, the ridge shape $h(x,t)=u_y(x,t)$ follows purely from the normal stress $\vec \sigma \cdot \vec e_y$, so that (\ref{eq:greensystem2}) combined with (\ref{eq:delta}) becomes

\begin{equation}\label{eq:hieps}
K_{y\phi}\otimes \phi = \phi' \Upsilon_s + \gamma \sin \theta \delta(x).
\end{equation}
Here, the discontinuity in $\phi$ at the contact line is imposed by means of a $\delta$ function. Since $\Upsilon_s$ is assumed constant across the entire interface, we can solve this equation by a spatial Fourier transform, using the travelling wave description, so that 

\begin{eqnarray}
\frac{\mu(qv)}{k_{y\phi}(q)} \tilde \phi_c(q)  = iq \Upsilon_s \tilde \phi_c +  \gamma \sin \theta.
\end{eqnarray}
Using that $\phi_c(x) = \partial h_c/\partial x$, this can be solved as

\begin{eqnarray}\label{eq:shapesolution}
\tilde h_c (q) = \frac{\tilde \phi_c(q)}{iq}  = \frac{\gamma \sin \theta}{\Upsilon_s q^2 + \frac{\mu(qv)}{k(q)} }.
\end{eqnarray}
Here we furthermore introduced $k(q)=k_{y\phi}(q)/(iq)$ as the more familiar Green's function that relates normal traction to normal displacement. For an incompressible layer of thickness $h_0$ this is given by 

\begin{equation}
  \label{eq:green}
k(q)=\left[\frac{\sinh(2qh_0) - 2qh_0}{\cosh(2qh_0) + 2(qh_0)^2+1 }\right]\;\frac{1}{2q},
\end{equation}
so that the ridge shape can be computed, for arbitrary substrate thickness. The velocity dependence is encoded through the argument of the complex modulus in (\ref{eq:shapesolution}).

\subsection{Neumann's law and ridge rotation}

The analytical solution (\ref{eq:shapesolution}) has two important features. First, we verify that we recover the boundary condition in the form of a discontinuity of the slope at the contact line. This follows from the large-$q$ asymptotics of (\ref{eq:shapesolution}). Noting that $k(q)\sim 1/|q|$ and assuming that at large frequency $\mu(\omega) \ll \omega$, we find the dominant behaviour $\widetilde{h}_c \sim \gamma/(\Upsilon_s q^{2})$, which indeed implies a slope-discontinuity of strength $\gamma/\Upsilon_s$. In the limit of small slopes, this implies a solid angle

\begin{equation}\label{eq:verticalneumann}
\theta_S = \pi - \frac{\gamma \sin \theta}{\Upsilon_s}.
\end{equation}
Reminding that the formalism a priori assumed small substrate deformations, i.e. $\gamma/\Upsilon_s \ll 1$, the result for $\theta_S$ can be indeed recognised as the vertical component of Neumann's law. The value of $\theta_S$ depends only on the surface tensions, and is totally independent of $\mu(\omega)$ and of the contact line velocity. From the asymptotics it is clear this is satisfied whenever $\mu(\omega) \ll \omega$ at large frequency. This is indeed the case when $\mu \sim \omega^n$ with $n<1$ at large frequency. This confirms the condition of integrable stress, as discussed in  Sec.~\ref{subsec:neumann}. 

A second important feature of the ridge solution (\ref{eq:shapesolution}) is that due to the motion the shape becomes asymmetric, and the ridge tip exhibits a rotation that we describe by an angle $\varphi$. The rotation can be calculated by taking the symmetric (real) part of $h'$: 
\begin{align}
\label{eq:tanphi}
 \varphi &\approx \lim_{x \rightarrow 0}\frac{1}{2}(h'(x)+h'(-x))  =\frac{1}{2\pi} \int \Re[-iq\tilde{h}_c(q)]e^{-iqx}dq .
\end{align}
Using the explicit form (\ref{eq:shapesolution}), the rotation angle becomes

\begin{equation}
\label{eq:neumann-rotation}
\varphi = \gamma \sin \theta \int \frac{dq}{2\pi} \frac{qk(q)G''(qv)}{\mid k(q)\Upsilon_s q^2 + \mu(vq)\mid^2}.
\end{equation}
Its small $v$ asymptotics gives (\ref{eq:smallspeed}), as derived in~\cite{KarpNcom15}. One verifies that the expression for (\ref{eq:neumann-rotation}) is integrable, if and only if $n<1$; so the same condition on the rheology appears once more.

\subsection{Consistency check: Dissipation approach.}\label{sec:dissipation}
~As a consistency check, we now perform an alternative calculation of the liquid contact angle. We follow the dissipation approach as originally proposed by Long, Ajdari \& Leibler~\cite{LALLang96}, and reused recently by Zhao \emph{et al.}\cite{zhao2018geometrical}. The analysis builds on a balance of the work per unit time performed by the capillary force and the dissipation inside the layer, i.e. 
\begin{equation}
\label{eq:powerinject}
P=\gamma v (\cos\theta_{eq}-\cos\theta),
\end{equation}
where $P$ is the total dissipation (per unit contact line length)
\begin{equation}
\label{eq:dissipation}
P =\int d^2x \, \boldsymbol{\sigma} :   \nabla \dot{\vec u}.
\end{equation}

The integral of (\ref{eq:dissipation}) can be brought to the free surface using Gauss's divergence theorem

\begin{equation}
\label{eq:dissipationgauss}
P =\int d^2x \, \boldsymbol{\sigma} :   \nabla \dot{\vec u} = 
\oint ds \,    \vec \sigma \cdot \dot{\vec u}.
\end{equation}
A justification of the use of the divergence theorem will be given in Appendix~\ref{app:gauss}. Given that the displacement vanishes at the bottom of the substrate, the only contribution comes from the integral over the free surface, where for small deformations the normal displacement reads $h(x,t)$. Without the Shuttleworth effect, the traction only has a normal component, so that

\begin{equation}
\label{eq:dissipationtres}
P =
\int_{-\infty}^\infty dx \, \sigma_{yy} (x,t)  \dot{h}(x,t),
\end{equation}
In the frame comoving with the contact line, the elastic stress at the free surface can be computed from the Green's function,
\begin{equation}
\tilde \sigma_{c,yy}(q) = \frac{\mu(qv)}{k(q)} \tilde h_c(q),
\end{equation}
and $\tilde{\dot h} = iqv\, \tilde h_c(q)$. 
Then, (\ref{eq:dissipationtres}) can be written as 

\begin{align}
\label{eq:fourierdissipation}
P &= \int d\bar x \left\{ \int \frac{dq}{2\pi}  \tilde \sigma_{c,yy}(q) e^{iq \bar x}\right\}
\left\{ \int \frac{dq'}{2\pi}   iq'v\, \tilde h_c(q') e^{iq' \bar x} \right\} \nonumber \\ 
&= \int d\bar x  
\int \frac{dq}{2\pi} \int \frac{dq'}{2\pi}  \frac{iq'v\, \mu(qv)}{k(q)} \tilde h_c(q)  
\tilde h_c(q') e^{i(q+q') \bar x}, 
\end{align}
where we introduced the change of variables $\bar x=x-vt$. Using the identity $\int d\bar x \, e^{i(q'+q)\bar x}=2\pi \delta(q+q')$, we finally obtain

\begin{align}
\label{eq:dissipation_layer}
P &= - v \int \frac{dq}{2\pi}  \frac{\mu(qv)}{k(q)} iq\, |\tilde h_c(q)|^2 \nonumber \\
&=  
 v \int \frac{dq}{2\pi}  \frac{q G''(qv)}{k(q)} |\tilde h_c(q)|^2.
\end{align}
This expression provides the dissipation for any travelling wave of shape $h_c(x)$, regardless of the traction that generates it. Naturally, the dissipation is strictly positive and only involves the loss modulus $G''$. The final step is to impose the solution for the shape (\ref{eq:shapesolution}), so that the dissipation becomes

\begin{equation}
\label{eq:dissipation_layerbis}
P = v \left(\gamma \sin \theta \right)^2 \int \frac{dq}{2\pi}  \frac{q k(q)G''(qv)}{ |k(q)\Upsilon_s q^2 +  \mu(vq)|^2 }.
\end{equation}
For small changes in the contact angle, $\Delta \theta = \theta - \theta_{eq} \ll 1$, the dynamic contact angle selection (\ref{eq:powerinject}) then gives the final result

\begin{equation}
\label{eq:rotation_by_dissipation}
\Delta \theta = \gamma \sin \theta \int \frac{dq}{2\pi}  \frac{q k(q)G''(qv)}{ |k(q)\Upsilon_s q^2 +  \mu(vq)|^2 }.
\end{equation}

Hence, the dissipation approach gives a closed form expression for the dynamic contact angle of the liquid, for arbitrary rheology and arbitrary layer thickness -- and without any a priori assumptions on Neumann's law. One verifies that the expression (\ref{eq:rotation_by_dissipation}) for $\Delta \theta$ is indeed strictly identical to the expression for ridge rotation $\varphi$ obtained in (\ref{eq:neumann-rotation}); as is also observed experimentally at small velocity. This confirms once more $\varphi= \Delta \theta$ and hence the validity of Neumann's law. As a final remark, we note that the dissipation expression (\ref{eq:dissipation_layerbis}) is integrable only for $n<1$, so the same condition on the rheology appears yet again.

\section{On the validity of using the divergence theorem to estimate dissipation}\label{app:gauss}

Here we motivate the validity of using the divergence theorem (\ref{eq:dissipationgauss}) in the case where the interface slope exhibits a discontinuity at the contact line. To this end, we replace the Dirac $\delta$-function in (\ref{eq:hieps}) by a smooth function of width $a$, i.e. 

\begin{equation}\label{eq:Tractionsmooth}
T(x,t)= \gamma \sin \theta \frac{1}{a}f\left(\frac{x-vt}{a} \right).
\end{equation}
For example, one could consider $f$ to be a Gaussian, which in the limit of $a\rightarrow 0$ gives a representation of the Dirac $\delta$ function -- in fact, the true capillary traction is not infinitely sharp, but has $a$ of the order of the nanometric width of the interface~\cite{White:2003aa,Lub14,Weijs2013a,MDSA12b}. Carrying through this modification, one ends up with smooth, differentiable stress and displacement fields, regularising the slope discontinuity at the contact line. Hence, with (\ref{eq:Tractionsmooth}) there is no uncertainty in the use of Gauss's divergence theorem. It gives 

\begin{equation}
\label{eq:rotation_by_dissipationsmooth}
\Delta \theta_a = \gamma \sin \theta \int \frac{dq}{2\pi}  \frac{q k(q)G''(qv) |\tilde f(qa)|^2}{ |k(q)\Upsilon_s q^2 +  \mu(vq)|^2 }.
\end{equation}
The factor $\tilde f \rightarrow 1$ in the limit of $|qa| \ll 1$, so that the integrand of (\ref{eq:rotation_by_dissipationsmooth}) approaches the result (\ref{eq:rotation_by_dissipation}) as $a \rightarrow 0$. To ensure that this limit is not singular, we also evaluate the integral at finite $a$, and consider the limit of vanishing $a$ after integration. Clearly, $\tilde f(qa)$ acts as a cutoff of the integral beyond wavenumbers $q \sim a^{-1}$. Since the integrand scales as $\sim q^{n-2}$, the integral for small but finite $a$ will scale as $\sim (a^{-1})^{n-1}\sim a^{1-n}$, which is convergent in the limit $a\rightarrow 0$ as long as $n<1$. 

In summary, $\Delta \theta_a$ for a ``smoothened" contact line of finite width $a$ will converge to $\Delta \theta$ as computed in (\ref{eq:neumann-rotation}) in the limit $a\rightarrow 0$. The only requirement for the analysis to be valid is that the dissipation $P$ is integrable when $a\rightarrow 0$, which requires $n<1$.

\bibliographystyle{apsrev4-1}

%

\end{document}